\documentclass[aps,prd,reprint,twocolumn,preprintnumbers,floatfix,nofootinbib]{revtex4-1}

\pdfoutput=1

\usepackage{graphicx}
\usepackage{epsfig}
\usepackage{amsmath}
\usepackage{empheq}
\usepackage{amsfonts}
\usepackage{amssymb}
\usepackage{color}
\usepackage[bottom]{footmisc}
\usepackage{array,multirow,makecell}
\usepackage{slashed}
\usepackage[colorlinks,citecolor=blue]{hyperref}
\usepackage{float}
\usepackage{ulem}
\usepackage[utf8]{inputenc}
\usepackage[english]{babel}
\usepackage{url}
\usepackage{hyperref}
\usepackage{xcolor}
\usepackage{subfigure}
\usepackage{lipsum}
\usepackage[bottom]{footmisc}
\usepackage{pifont}

\usepackage[most]{tcolorbox}
\newtcbox{\mymath}[1][]{%
    nobeforeafter, math upper, tcbox raise base,
    enhanced, colframe=blue!30!black,
    colback=blue!30, boxrule=1pt,
    #1}

\usepackage{blindtext}

\def\lsim{\raise0.3ex\hbox{$\;<$\kern-0.75em\raise-1.1ex\hbox{$\sim\;$}}}
\def\gsim{\raise0.3ex\hbox{$\;>$\kern-0.75em\raise-1.1ex\hbox{$\sim\;$}}}

\newcommand{\bea}{\begin{aligned}}
\newcommand{\eea}{\end{aligned}}
\def\beq{\begin{equation}}
\def\eeq{\end{equation}}
\def\beqa{\begin{eqnarray}}
\def\eeqa{\end{eqnarray}}
\def\be{\begin{equation}}
\def\ee{\end{equation}}

\def\bse{\begin{subequations}}
\def\ese{\end{subequations}}

\def\trh{$T_{\mathrm{RH}}~$}
\def\mrm{\mathrm}
\def\tin{t_{\rm in}}
\def\tev{t_{\rm ev}}
\def\Gammaev{\Gamma_{\rm ev}}
\def\trh{T_{\mathrm{RH}}}
\def\ain{a_{\rm in}}
\def\ae{a_{\mathrm{end}}}
\def\Hend{H_{\rm end}}
\def\He{H_{\rm end}}
\def\arh{a_{\mathrm{RH}}}
\def\abh{a_{\mathrm{BH}}}

\def\rhobh{\rho_{\rm BH}}
\def\ahc{a_{\rm hc}}
\def\khc{k_{\rm hc}}
\def\rhorh{\rho_{\mathrm{RH}}}
\def\krh{k_{\rm RH}}
\def\rhogw{\rho_{\rm GW}}
\def\bea{\begin{eqnarray}}
\def\eea{\end{eqnarray}}
\def\rhoe{\rho_{\mathrm{end}}}
\def\aend{a_{\rm end}}
\def\omegae{\omega_{\rm end}}
\def\ai{a_{\mathrm{in}}}
\def\aev{a_{\mathrm{ev}}}

\def\Min{M_{\mathrm{in}}}

\def\kin{k_{\rm in}}

\def\mbh{M_{\rm BH}}

\def\Tbh{T_{\rm BH}}
\def\gammabh{\Gamma_{\rm BH}}
\def\tq{t_q}
\def\Trh{T_{\rm RH}}
\def\nbh{n_{\rm BH}}
\def\dbh{d_{\rm BH}}

\newcommand{\baz}{\begin{array}{cc}}

\newcommand{\bav}{\begin{array}{cccc}}

\newcommand{\noi}{\noindent}

\begin{document}

 \title{
 Burdening (or not) gravitational waves in the presence of primordial black holes}

\author{Mathieu Gross}
\email{mathieu.gross@ijclab.in2p3.fr}
\affiliation{Universit\'e Paris-Saclay, CNRS/IN2P3, IJCLab, 91405 Orsay, France}
\author{Yann Mambrini}
\email{yann.mambrini@ijclab.in2p3.fr}
\affiliation{Universit\'e Paris-Saclay, CNRS/IN2P3, IJCLab, 91405 Orsay, France}
\author{Md Riajul Haque}%
\email{riaj.0009@gmail.com}
\affiliation{\,Physics and Applied Mathematics Unit, Indian Statistical Institute, 203 B.T. Road, Kolkata 700108, India}.

\begin{abstract}

We present the spectrum of primordial gravitational wave (GW) expected from the presence of primordial black holes (PBH) and inflaton in the early Universe. For the first time, we combine the waves produced by the PBH decay, with their density fluctuation  counterpart, as well as their effects on the GW produced by the inflaton {\it after} (high frequency modes) and {\it before} (low frequency modes) the end of inflation. We generalize our study for a potential $V(\phi)\propto \phi^k$ during reheating. We also extend our study, taking into account a possible memory burden effect to see how it can affect the shape of the spectrum.

\end{abstract}

\maketitle


\section{Introduction and conclusion}

The discovery of gravitational waves (GW) at the Hertz scale 
in 2015 \cite{LIGOScientific:2016aoc,LIGOScientific:2016dsl,LIGOScientific:2016wyt} opened a new window in the physics of black holes
\cite{LIGOScientific:2017bnn,Abbott_2021,LIGOScientific:2017ycc,LIGOScientific:2017vox}. 
Several experiments like the Einstein Telescope \cite{Hild:2010id} or e-LISA \cite{LISA:2017pwj} proposed to cover a larger range of frequencies, down to 
the micro-Hertz, whereas pulsar timing array already analyzed secondary GW generated by hypothetical PBH
down to the nanohertz \cite{NANOGrav:2023gor,NANOGrav:2023hde,EPTA:2023sfo,EPTA:2023fyk,EPTA:2023fyk,Zic:2023gta,Reardon:2023gzh}. However, another type of gravitational 
waves is also of great interest to study the cosmology in early 
Universe, at larger frequencies. Indeed, different gravitational sources exists from the inflationary era to the BBN epoch.

\begin{figure}[!h]
\centering
\vskip .2in
\includegraphics[width = 0.5\textwidth]{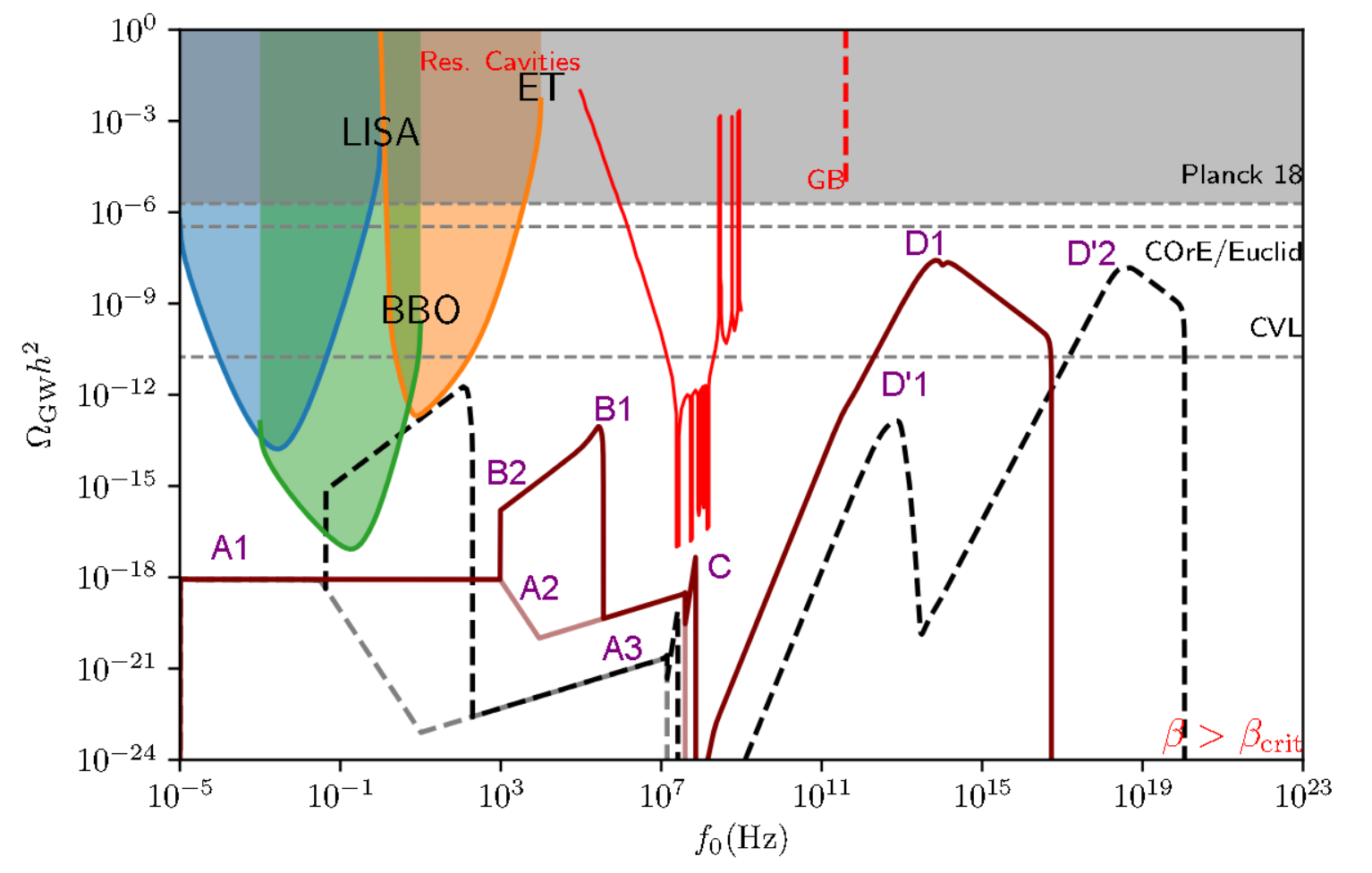}
\caption{
 Minimal Gravitational wave spectra coming from a PBH reheating phase after inflation, with an inflaton potential $V(\phi)\propto \phi^6$ during reheating. The thick red curve corresponds to $\beta = 10^{-5}$ while the dashed black one takes care of the memory burden effect, Eq.~(\ref{Eq:hawkingburden}), with $n=1$ and $\beta = 10^{-8}$. For both curves $\mbh= 1~g$. The regions $A_i$, $B_i$, $C$ and $D_i$ are described in the text.
}
\label{Fig:masterplot}
\end{figure}

The high frequency GW spectrum we obtained,
taking into account the two main sources in  the early Universe (inflaton and PBH) is represented in
Fig.~\ref{Fig:masterplot}.
It shows
the relic density observed today, $\Omega_{GW}h^2$ as function 
of the present frequency $f_0$. As an illustration, this spectrum was obtained for a population of 
1 gram PBHs, responsible for the reheating in the Universe \cite{RiajulHaque:2023cqe}. We incorporate a possible 
memory burden effect \cite{Alexandre:2024nuo,Dvali:2024hsb,Thoss:2024hsr} through the light dashed line. 
 We also display some sensitivity prospects for future gravitational wave observatories such as BBO~\cite{Crowder:2005nr,Corbin:2005ny}, LISA~\cite{LISA:2017pwj} and Einstein telescope (ET)~\cite{ET:2019dnz,Sathyaprakash:2012jk,Abac:2025saz} as well as current and projected experimental constraints from CMB measurements (Planck, COrE/Eu-
clid and CVL) \cite{COrE:2011bfs,EUCLID:2011zbd,Ben-Dayan:2019gll}.

Our paper is organized to explain 
the features of the GW spectrum, from the lower frequencies to the highest ones. 
On the far left, the flat part of the spectrum labeled $A_1$, corresponds to the primordial GW modes which 
have entered in the horizon {\it after} the end of reheating, {\it ie.} during the radiation 
dominated era. After reminding the basis of PBH physics, and the subtleties of the burden 
effect in section II, we explain this range of frequencies in section III$_{\rm A}$. 

Following the red (full) line, we then observe a peak between $f_0\simeq 1-100$ kHz,
labeled $B_2$ and $B_1$. These GW are generated by the density fluctuations of 
the PBH background. The peak frequency corresponds roughly to the 
mean separation between PBHs at evaporation time, whereas the lower one to the size of the Universe at the evaporation time. 
This mechanism is described in section V. 

After the peak, the density of GW increases as 
a power law with the frequency in the region $A_3$, between $\sim 10^5$ and $\sim 10^7$ Hz. These GW are the modes entering in the horizon {\it before} the PBHs decay, when 
the Universe is still dominated by the oscillations of the inflaton field $\phi$, whose equation of state is $P_\phi=w_\phi\rho_\phi$. We took $w_\phi=\frac12$ for the illustration, which we explain in section III$_{\rm C}$. 

The brutal cut in the spectrum at $f_0\simeq 10^7$ Hz, region $C$, 
corresponds to the modes which never exited the horizon during inflation. They
can be interpreted as sourced by prompt inflaton scattering, as we explain in section IV.

Finally, around $f_0\simeq 10^{15}$ Hz, we 
observe a (double) peak labeled $D_1$ ($D_1'$ and $D_2'$) which is generated by 
the gravitons directly produced through the evaporation of the (burdened) PBH. This happens at 
frequencies corresponding roughly to the redshifted
Hawking temperature, and is explained in section VI. We then conclude in section VII.

\section{The physics of PBH}

The presence of primordial black holes can drastically 
change the physics in the early Universe, particularly with regard to GW. 
They can play 
a role in the expansion rate if they come to dominate over the inflaton field \cite{Haque:2023awl}, be the main source the reheating 
process if they lived sufficiently long enough \cite{RiajulHaque:2023cqe},
and even populate the dark component \cite{Haque:2023awl} or induce leptogenesis \cite{Barman:2024slw}. They 
also have the possibility to 
generate gravitational wave \cite{Gross:2024wkl,Paul:2025kdd,Balaji:2024hpu,Qiu:2025tmn,Ai:2025fqw}, or modify their spectrum. See \cite{Gunn:2025zry} and \cite{Kohri:2025avm}
for recent reviews on the subject.
All these effects depend on the evaporation process, 
which is also up for discussion, especially considering memory burden corrections to the semi-classical approach \cite{Haque:2024eyh}.

\subsection{The semi-classical approach}

From the seminal work of S Hawking in 1974 \cite{Hawking:1974rv}, it is known that 
a black hole can "explode", releasing 
its entropy under the form of the Hawking radiation \cite{Hawking:1975vcx}, described by the equation

\begin{equation}
\frac{d \mbh}{dt}=-\epsilon \frac{M_P^4}{\mbh^2}\,,
\label{Eq:hawking}
\end{equation}

\noi
with 
$\epsilon =\frac{27}{4}\frac{\pi g_{\star}(\Tbh)}{480}$,
where
$g_{\ast}(\Tbh)$ is the number of degrees of freedom 
associated with the PBH temperature, and 
$M_P = 1/\sqrt{8\pi G} \simeq 2.4 \times 10^{18} \, \rm{GeV}$ is the reduced Planck mass. 
The factor ${27}/{4}$~\cite{MacGibbon:1990zk} accounts for the greybody 
factor\footnote{Which can be more complex 
depending on the context \cite{Auffinger:2020afu,Masina:2021zpu,Cheek:2021odj,Arbey:2025dnc}. Note that we neglected the 
accretion effect \cite{Das:2025vts,Maity:2025ffa}.}. This energy loss corresponds to the Hawking temperature

\beq
\Tbh=\frac{M_P^2}{\mbh}\simeq 10^{13}\left(\frac{1~\rm g}{\mbh}\right)~{\rm GeV}\,.
\label{Eq:thawking}
\eeq

Solving Eq.~(\ref{Eq:hawking}) leads to the time evolution of the PBH mass

\beq
\mbh(t)=\Min\left(1-\frac{t-\tin}{\tev}\right)^\frac13,
\label{Eq:mbht}
\eeq
where $t_{\rm in}$ is the time at formation, and $\tev$ is the evaporation time, i.e.

\beq
\tev =\gammabh^{-1}=\frac{\Min^3}{3\epsilon M_P^4}
\simeq 2.4\times 10^{-28} \left(\frac{\Min}{1~\rm{g}}\right)^3~\rm{s}\,.
\label{Eq:tev}
\eeq
We see that PBHs with masses $\lesssim 10^9$g decay {\it before}
the beginning of the nucleosynthesis process, and are not constrained by CMB data. 
On the other hand, the initial PBH mass $M_{\rm in}$ is bounded by the size of the horizon at the end of inflation,
\bea
\Min \gtrsim H_{\rm end}^{-3} \rho_{\rm end} \sim \frac{M_P^3}{\sqrt{\rho_{\rm end}}}\simeq 1~\mrm{g}\,.
\eea
Taking into account the above constraints, we restrict our analysis to the following PBH mass range: 
\bea
1 \mrm{g} \lesssim \Min \lesssim 10^9 ~\mrm{g}.
\eea
However, this mass range drastically changes if one takes into account the memory burden effect, as we will see in a following section.

\subsection{PBH reheating}

It was shown 
in \cite{RiajulHaque:2023cqe,Haque:2023awl}
that PBHs can reheat the Universe through their Hawking radiation, even in the presence of an inflaton field, 
whose oscillations dominate the energy budget of the Universe at the end of inflation. Obviously, if the PBHs lifetime is longer than the inflaton one, the PBH population dominates the reheating process. However, as it was shown in \cite{RiajulHaque:2023cqe}, this is not the only situation where the reheating is dominated by the PBHs. Indeed, as the inflaton 
energy density $\rho_\phi\propto a^{-3(1+w_\phi)}$, with an equation of 
state $P_\phi=w_\phi \rho_\phi$, for $w_\phi>0$, there exists the possibility
for the PBHs to dominate the energy budget even {\it before} the inflaton decay. This happens for a sufficient amount of PBH at formation time. 
Defining 

\beq
\beta=
\frac{\rhobh(\tin)}{\rho_{tot}(\tin)}
\simeq \frac{\rhobh(\tin)}{\rho_\phi(\tin)}\,,
\eeq
it was shown in \cite{RiajulHaque:2023cqe} that PBH energy density dominates over the inflaton energy density for $\beta$
above a critical value

\begin{eqnarray}
\label{Eq:betamin}
\beta_{\rm c} & = & \left(\frac{\epsilon}{(1+w_\phi)2 \pi \gamma}\right)^{\frac{2w_\phi}{1+w_\phi}} \left(\frac{M_P}{\Min}\right)^{\frac{4w_\phi}{1+w_\phi}} \,,
\end{eqnarray}
where $\gamma$ is the collapse efficiency parameter, of the order ${\cal O}(1)$,
defined by

\begin{equation}
M_{\rm in} = \frac{4\pi\gamma M_p^2}{H_{\rm in}}, \quad \gamma = w_\phi^{3/2}\,,
\end{equation}
see appendix \ref{Sec:pbhdecay}, and especially Eq.~(\ref{Eq:betacburden}) for
a details computation of $\beta_c$.
This corresponds for instance to {$\beta_c \simeq 3\times 10^{-6}$ } for a quartic potential of the inflaton, $V(\phi)\propto \phi^4$ ($w_\phi=1/3$), and $\Min = M_{\rm min} = 1$ g.

In the present work, we consider that the reheating is induced by the PBHs population. In other words, the
successive phases are : inflation, inflaton oscillations, PBH domination, 
followed by a radiative Universe. In this case, 
the reheating temperature is obtained by solving

\beq
\rhorh=\alpha \Trh^4 = 3 M_P^2 H^2(\Trh)=\frac{4}{3}\frac{M_P^2}{\tev^2}=12\epsilon^2\frac{M_P^{10}}{\mbh^6}\,,
\label{Eq:rhorhpbh}
\eeq

\noi
which implies
\beq
\Trh=\left(\frac{12 \epsilon^2}{\alpha}\right)^\frac14~\sqrt{\frac{M_P^5}{\mbh^3}}
\simeq 3.5\times 10^{10}\left(\frac{1~\rm g}{\mbh}\right)^\frac32 \rm GeV \,,
\label{Eq:trh}
\eeq

\noi
with $\alpha=\frac{g_{RH} \pi^2}{30}$, $g_{RH}$
being
the relativistic degrees of freedom at $\Trh$. Note that,
in the semi-classical approach, we can approximate
$\mbh\simeq \Min$ during the whole evaporation process.
This will not be the case when we examine the possibility of a memory burden effect.  
Note that $\trh$ is different from the Hawking temperature $\Tbh$ in Eq.~(\ref{Eq:thawking}), even if they are 
related in someway. The Hawking temperature represents 
the {\it efficiency} of the evaporation, whereas the reheating 
temperature represents the {\it lifetime} of the PBH, and is 
then naturally cooler.

\subsection{The burden effect}
\label{Sec:burden}

It was shown that in a system of high storage 
capacity, the information stored in the system tends to backreact to resist to the process of 
information lost \cite{Dvali:2018xpy,Dvali:2018ytn}. Applied to black holes, this corresponds to a 
modification of the semi-classical approach to their decay process \cite{Alexandre:2024nuo,Dvali:2024hsb,Thoss:2024hsr}. 
Naively, this backreaction effects should 
become efficient
when the leftover information is of the 
magnitude of the information lost, or $\Delta \mbh\sim \mbh$.
The authors of \cite{Dvali:2025ktz} even proposed
that the burdening sequence is a 
continuous process, beginning at the 
onset of their formation. The aim of this 
section is to explain with simple arguments how the memory burden affects the phenomenology of PBH in the early Universe. For interested readers, we've devoted a complete section \ref{Sec:pbhdecay} in the appendix, 
where all our results are derived in detail. 

When we take into account the memory burden effect, the evaporation of a PBH occurs in 
two phases. Initially, the PBH behaves in a semi-classical way, until a non-negligible 
fraction of lost information triggers a black hole back-reaction, effectively slowing 
down the evaporation process. To model the phenomenon, we 
suppose that the semi--classical regime is valid until the mass of the PBH reaches a certain value,
$\mbh=q \Min$, with $0<q<1$.
There are no real microscopic motivated values for $q$, but we can expect that when $q\sim \frac12$, the amount of information lost is of the order of the amount of 
information still inside the PBH. At that stage, the back-reaction effect should be 
triggered, as a survival reflex of the PBH. This value of $q$ is also the benchmark 
point used by the authors of \cite{Alexandre:2024nuo,Thoss:2024hsr}.

The same authors proposed to model the back-reaction effect, by a change in the 
semi-classical evaporation equation,

\beq
\left.\frac{d\mbh}{dt}\right|_{burden}=\frac{1}{S^n}\frac{d\mbh}{dt}
=-2^n \epsilon \times \frac{M_P^{2n+4}}{\mbh^{2n+2}}\,,
\label{Eq:hawkingburden}
\eeq
where $d\mbh /dt $ is given by Eq.~(\ref{Eq:hawking}) and $n>0$. In Eq.~(\ref{Eq:hawkingburden}), $S$ is the PBH entropy

\beq
S=\frac12\frac{\mbh^2}{M_P^2}\,.
\eeq

\noi
Even if the dependence on $S$ in Eq.~(\ref{Eq:hawkingburden}) can be justified by analogy with the theory
of information \cite{Dvali:2018xpy,Dvali:2018ytn,Dvali:2020wft}, we can also stay agnostic and suppose that the loss in the 
decay rate is an inverse function of the entropy, which can be modeled by a power law at the first order. The evolution of the PBH is then modified, and is obtained
by integrating Eq.~(\ref{Eq:hawkingburden}) :

\beq
\mbh(t)=q\Min\left[1-\gammabh^n(t-\tq)\right]^{\frac{1}{3+2n}}\,,
\label{Eq:mass_burden}
\eeq
with
\beq
\gammabh^n= {2^n(3+2n)\,\epsilon}\times M_P
\left(\frac{M_P}{q\Min}\right)^{3+2n}\,.
\label{Eq:mass2}
\eeq 

\noi
and

\bea 
\tq= \frac{1-q^3}{\gammabh^0},
\label{Eq:tq}
\eea 
is the time which ends the semi-classical phase. Note that for $n=0$ and $q=1$, we recover the semiclassical PBH lifetime $(\Gamma_{\rm BH})^{-1}=\frac{\Min^3}{3 \epsilon M_P^4}$.

The reheating temperature obtained in Eq.~(\ref{Eq:trh}) then becomes 

\bea\label{eq: tempmemory}
&&
\alpha \Trh^4=\frac{4}{3}
M_P^2 (\Gamma^n_{\rm BH})^2=
\frac{4\times 2^{2n}(3+2n)^2}{3 q^{6+4n}}
\epsilon^2\frac{M_P^{10+4n}}{\Min^{6+4n}}
\nonumber
\\
&&
\Rightarrow
\left.\Trh\right|_{n=1}^{q=\frac12}
\simeq 1.6\times 10^6~\left(\frac{1~\rm g}{\Min}\right)^{\frac52}~\rm GeV\,,
\label{Eq:trhburden}
\eea

\noi
which is lower than the reheating temperature {\it without} burdening, 
as expected because of a longer PBH lifetime. Other constraints on the burden parameters can be inferred from BBN bounds \cite{Chaudhuri:2025asm}, WIMP/FIMP production modes \cite{Takeshita:2025mhy}, or observations of galactic 
and extra-galactic sources \cite{Tan:2025vxp,Chianese:2025wrk,Athron:2024fcj,Chianese:2024rsn,Saha:2024ies} and lepto/baryogenesis \cite{Calabrese:2025sfh,Borah:2024bcr}.

\section{Primordial gravitational wave}\label{sec: 3}

It is well known that quantum fluctuations are sources of gravitational waves during inflation \cite{Guzzetti:2016mkm}.
These stochastic gravitational waves are called {\it primordial} gravitational waves (PGW). 
Because the gravitons do not thermalize, PGWs are a unique opportunity to test the physics of reheating. 
Given the energy density 
involved during inflation,
the spectrum 
of gravitational waves produced 
covers a very large range of frequencies. 
For instance, at the end of inflation, the energy available under the form of oscillations of the inflaton, is of the order of $\sim m_\phi$.
It would generate waves of frequency

\beq
f_{\rm end}\sim \left(\frac{m_\phi}{4\times 10^{13}~\rm GeV}\right)10^{37}~{\rm Hz}\,, 
\eeq
which redshifts as $\aend/a_0$ nowadays. The redshift is obviously highly dependent on the species which dominate the expansion rate $H$, and the reheating process. However, in a radiation-like Universe, which happens in the case of an instant reheating, or a Universe dominated by the oscillations of an inflaton with a quartic potential, we have $\aend/a_0\sim T_0/\rhoe^\frac14 \sim 10^{-29}$, where we took $\rhoe \sim 10^{64}$ GeV$^4$, or

\beq
f_0^{\rm end}\sim m_\phi\frac{\aend}{a_0}\sim \left(\frac{m_\phi}{4\times 10^{13}~\rm GeV}\right)10^{8}~\rm Hz\,.
\label{Eq:fend}
\eeq
This frequency corresponds roughly to the end of the PGW spectrum, region {\bf $A_3$/$C$}
in Fig.~\ref{Fig:masterplot}.
In the following, we will describe the energy spectrum from the lower frequencies
(region $A_1$ of Fig.~\ref{Fig:masterplot}, corresponding to the modes entering in the horizon during the radiation dominated era)
to the higher ones (region $A_3$, corresponding to the modes entering at the very end of inflation).

\subsection{The radiation dominated era ($A_1$)}

The density of energy stored under the form of GW of mode $k$ can be estimated by \cite{Saikawa:2018rcs,Chakraborty:2025zgx}

\beq
\rho_{GW} \equiv \frac{d \rho_{GW}(k)}{d \log k} \sim k^2 \langle h_{ij}h_{ij}\rangle\,,
\label{Eq:pgw}
\eeq

where 

\beq
\langle h_{ij}h_{ij} \rangle\sim \left(\frac{H_{\rm end}}{2\pi}\right)^2\,,
\eeq
are the tensor fluctuations produced during inflation.
The expression (\ref{Eq:pgw}) can be understood noting that the solution
of the equation of motion for modes $h_{ij}$ staying below the horizon
during the whole inflation ($k \ll \He$) behaves as \cite{Ema:2020ggo}

\beq
h_{ij}\sim\frac{\He}{\sqrt{2} k^{\frac32}}e^{i k t}\,,
\eeq
which implies

\beq
d \rho_{GW}\sim \frac{k^3}{2 \pi^2}  \nabla |h_{ij}|^2 d \ln k \sim k^2
\left(\frac{\He}{2 \pi}\right)^2 d \ln k\,.
\eeq

For interested readers, we develop in appendix \ref{App:pgw}, especially Eq.~(\ref{Eq:pgwspectrum}), the details of the 
 calculation which lead to the exact PGW spectrum that we used for 
our numerical analysis. However, the expression (\ref{Eq:pgw}) 
remains a good approximation for 
understanding the mechanism, and the final shape of the spectrum that
we explain in detail in appendix \ref{App:pgw}.

After the end of inflation, the wave mode $k< \He$ reenters inside the horizon when
$k\sim H$, with a frequency $f=k/2 \pi$.
Once produced, the energy density $\rho_{GW}$ evolves in the same way as 
the radiation density $\rho_R$, i.e $\rho_{GW} \propto a^{-4}$. The ratio $\rho_{GW}/\rho_R$ is then constant (up to decoupled degrees of freedom) and is given by

\beq
\frac{\rho_{GW}}{\rho_R}(a)= \frac{k^2 H_{\rm end}^2}{4 \pi^2 \rho_R} \sim
\frac{1}{12 \pi^2}\frac{H_{\rm end}^2}{M_P^2}\simeq 5.3\times 10^{-14}\,,
\label{Eq:rhogw}
\eeq
where we used the fact that the Universe 
is dominated by $\rho_R=3 M_P^2H^2$ after the 
reheating phase, and $H=k$ at horizon 
re-entry. In Eq.~(\ref{Eq:rhogw}), we took $H_{\rm end}=2.5\times 10^{-6}M_P$, which 
will be our benchmark point throughout our 
present study. The spectrum is flat, as 
expected.

The relic abundance $\Omega_{GW}h^2$ is then

\beq
\Omega_{GW}h^2=\frac{\rho_{GW}}{\rho_{tot}}h^2=\frac{\rho_{GW}}{\rho_R}\Omega_R h^2 \simeq 4.1\times 10^{-5}~\frac{\rho_{GW}}{\rho_R}\,,
\label{Eq:omegah2}
\eeq
where we neglected the change in  the relativistic degrees of freedom
between the horizon crossing and present time.

\noi
Combining with Eq.~(\ref{Eq:rhogw}), 
we obtain

\beq
\boxed{
\Omega_{GW}^{A_1}h^2 \simeq \frac{4.1\times 10^{-5}}{12 \pi^2}
\left(\frac{\He}{M_P}\right)^2\simeq 2.2\times 10^{-18}\,,
\label{Eq:A1}
}
\eeq

\noi
which is effectively what we observe in the region $A_1$ of Fig.~\ref{Fig:masterplot}, where the decoupling of the degrees of freedom has been taken into account.

This region $A_1$ corresponds to the gravitational waves produced {\it after} reheating. We 
clearly recognize the scale-invariant form 
of the spectrum, whose frequencies span from the lower ones, the ones entered in the 
horizon at late time, to the highest one, the ones entered in the horizon at the reheating 
time\footnote{Throughout this work, we will use the notion of ‘time’ (before, after) indiscriminately with the notion of scale factor.} $\arh$. The form of the spectrum then changes for higher frequencies, region $A_2$. This is because before $\arh$, the 
evolution of the Universe is drastically 
different. It is dominated by the 
PBH which behaves like dust, and not radiation.  
We therefore expect the spectrum break to appear at a frequency $f_0^{\rm RH}$ corresponding to the end of reheating, or

\beq
f^{\rm RH}_0=\frac{k_{\rm RH}}{2 \pi}\frac{\arh}{a_0}=
\sqrt{\frac{\alpha}{3}}\frac{\Trh^2}{2 \pi M_P}\left(\frac{g_0}{g_{RH}}\right)^\frac13\frac{T_0}{\Trh}
\eeq
which gives
\beq
\boxed{
f^{\rm RH}_0=g_0^\frac13g_{RH}^\frac16 \frac{T_0 \Trh}{6 \sqrt{10}M_P}\simeq 946 \left(\frac{1~\rm g}{\mbh}\right)^\frac32~\rm Hz\,,
\label{Eq:frhnoburden}
}
\eeq
where in the last expression, we used Eq.~(\ref{Eq:trh}). This frequency of transition PBH domination $\rightarrow$
radiation domination is indeed the one we observe in Fig.~\ref{Fig:masterplot}.

\subsection{The PBH domination era ($A_2$)}

For frequencies $f^0$ larger than $f^0_{\rm RH}$, the PGW modes entered in the horizon during the phase of PBH domination. 
The evolution of the Hubble rate 
(and thus the horizon of re-entry) is then not governed by a radiation field, 
but by a dust, following 
$H(a)\propto a^{-\frac32}$. Using Eq.~(\ref{Eq:pgw}), the present GW energy density is then

\bea
&&
\rhogw (\arh)=\rhogw (\ahc)\left(\frac{\ahc}{\arh}\right)^4=\frac{\He^2}{4 \pi^2}\khc^2
\left(\frac{\ahc}{\arh}\right)^4
\nonumber
\\
&&
=\frac{\He^2}{4 \pi^2}k^2_{\rm RH}\left(\frac{\ahc}{\arh}\right)^2
\nonumber
\\
&&
\Rightarrow ~\frac{\rhogw(\arh)}{\rho_R(\arh)}=\frac{1}{36 \pi^2}
\frac{\He^2}{M_P^2}\frac{\rhorh}{M_P^2 \krh^2}\,,
\eea
where $\ahc$ is the scale factor at horizon crossing, $\khc=H(\ahc)$, and

\beq
3M_P^2 \khc^2=\rho(\ahc)=\rhorh\left(\frac{\arh}{\ahc}\right)^3
~\Rightarrow~\frac{\ahc}{\arh}=\frac{\rhorh}{3 M_P^2 k_{\rm RH}^2}
\,.
\nonumber
\eeq

\noi
where we used $\rho(a_k)=3 M_P^2 k^2$ at the horizon entry of the mode $k$. From
$\rhorh$ given by Eq.~(\ref{Eq:rhorhpbh}), we deduce, neglecting the change in number of degrees of freedom (see the appendix \ref{App:pgw} and Eq.~(\ref{Eq:pgwspectrum}) for a detailed calculation),

\beq
\frac{\rhogw (a_0)}{\rho_R(a_0)}\simeq\frac{\rhogw (\arh)}{\rho_R(\arh)}
\simeq\frac{1}{36 \pi^2}\frac{\He^2}{M_P^2}\frac{\alpha \Trh^2}{M_P^2}\frac{T_0^2}{k_0^2}\,,
\eeq
which implies

\begin{empheq}[box=\fbox]{align}
\Omega_{GW}^{A_2} h^2 & \simeq
\frac{4.1\times 10^{-5}\alpha}{36 \pi^2}\left(\frac{\He \Trh}{M_P^2}\right)^2
\left(\frac{T_0}{2 \pi f_0}\right)^2
\nonumber
\\
&\simeq 2.3\times 10^{-19}
\left(\frac{\Trh}{10^{10}~\rm GeV}\right)^2\left(\frac{10^3~\rm Hz}{f_0}\right)^2 \,,
\label{Eq:A2}
\end{empheq}

\noi
where we used Eq.~(\ref{Eq:omegah2}), and which is effectively what we observe in Fig.~\ref{Fig:masterplot}. Replacing $\Trh$ by its expression as function of $\mbh$, Eq.~(\ref{Eq:trh}), we then obtain in the semiclassical approximation

\bea
&&
\Omega_{GW}^{A_2}h^2\simeq\frac{4.1\times 10^{-5}\epsilon\sqrt{3 \alpha}}{18 \pi^2}
\left(\frac{\He^2 M_P}{\mbh^3}\right)
\left(\frac{T_0}{2 \pi f_0}\right)^2
\nonumber
\\
&&
\simeq 2.8\times 10^{-18}\left(\frac{1~\rm g}{\mbh}\right)^3\left(\frac{10^3~\rm Hz}{f_0}\right)^2\,.
\eea

\subsection{The inflaton domination era ($A_3$)}

The shape of the PGW spectrum described above is valid throughout the period of  
PBH domination, which begins at a scale factor $\abh$ defined by

\beq
\rhobh(\abh)=\rho_\phi(\abh)\,.
\label{Eq:rhobhabh}
\eeq

\noi
To express this condition as function of $\mbh$, one needs to introduce a new parameter, $\beta$,
which is the ratio between the energy density stored under the form of PBH and the total (inflaton) energy density at the time of PBH formation $\ai$,

\beq
\beta=\frac{\rhobh(\ai)}{\rho_\phi(\ai)}\,.
\label{Eq:beta}
\eeq
Combining Eqs.~(\ref{Eq:rhobhabh}) and (\ref{Eq:beta}) with $\rho_\phi(\abh)=\rho_\phi(\ai)\left(\frac{\ai}{\abh}\right)^{3(1+w_\phi)}$ and
$\rhobh(\abh)=\rhobh(\ai)\left(\frac{\ai}{\abh}\right)^3$, we found

\beq
\frac{\ai}{\abh}=\beta^{\frac{1}{3 w_\phi}}\,.
\eeq

\noi
The size of the horizon at $\abh$, which defines the highest frequency of the PBH domination  era is then

\beq
H(\abh)=H(\ain)\left(\frac{\ain}{\abh}\right)^\frac{3(1+w_\phi)}{2}
=4 \pi \frac{M_P^2}{\mbh}\beta^{\frac{1+w_\phi}{2w_\phi}}\,,
\label{Eq:hbh}
\eeq
where we used the approximation that all the mass inside the horizon collapses into PBH
at $\ai$, or\footnote{The exact expression depends on the efficiency of the collapse defined
by a parameter $\gamma$ which is function of $w_\phi$ and described in the appendix.}

\beq
\mbh \sim \Min=\frac{4}{3}\pi H^{-3}(\ai)\rho_\phi(\ai)=4 \pi \frac{M_P^2}{H(\ai)}\,.
\eeq

The present frequency corresponding to $k_{\rm BH}=H(\abh)$ is then

\bea
&&
f_0^{\rm BH}= \frac{k_{\rm BH}}{2 \pi}\frac{\abh}{\arh}\frac{\arh}{a_0}=
\frac{1}{2 \pi 3^\frac13}\frac{H^\frac13(\abh)\rhorh^\frac13}{M_P^\frac23}
\frac{T_0}{\Trh}\frac{g_0^\frac13}{g_{RH}^\frac13}
\nonumber
\\
&&
=\frac{1}{2\pi}\left(\frac{4 \pi}{3}\right)^\frac13\frac{\rhorh^\frac13}{\mbh^\frac13}\beta^{\frac{1+w_\phi}{6w_\phi}}\frac{T_0}{\Trh}\frac{g_0^\frac13}{g_{RH}^\frac13}
\label{Eq:f0bhnoburden}
\\
&&
\simeq 2.6\times 10^6 
\left(\frac{\trh}{10^{10}~\rm GeV}\right)^\frac13
\left(\frac{1~\rm g}{\mbh}\right)^\frac13 \beta^\frac{1+w_\phi}{6w_\phi}~\rm Hz
\,,
\nonumber
\eea
where we used 

\beq
\frac{\abh}{\arh}=\left(\frac{\rhorh}{\rho(\abh)}\right)^\frac13=
\frac{\rhorh^\frac13}{(3M_P^2H^2(\abh))^\frac13}\,.
\eeq

Replacing the expression (\ref{Eq:trh}) for $\trh$, and taking $w_\phi=\frac12$, 
we find

\beq
\boxed{
f_0^{BH}\simeq 1.2 \times 10^4 \left(\frac{1~\rm g}{\mbh}\right)^\frac56
\sqrt{\frac{\beta}{10^{-5}}}~\rm Hz\,,
\label{Eq:fbh}
}
\eeq
which is effectively the inflection point of the transition PBH-domination $\rightarrow$ inflaton-domination ($A_2 \rightarrow A_3$)
in Fig.~\ref{Fig:masterplot}. The GW density at this frequency is then given by 
Eq.(\ref{Eq:A2}) with $f_0=f_0^{BH}$, or

\beq
\Omega_{GW} h^2(f_0^{BH})\simeq 3.9\times 10^{-20}\left(\frac{10^{-5}}{\beta}\right)
\left(\frac{1~\rm g}{\mbh}\right)^\frac43\,.
\label{Eq:omegabh}
\eeq

For larger frequencies, $f_0>f_0^{BH}$, the PGW propagates in a medium dominated by 
the inflaton field, whose equation of state is $p_\phi=w_\phi \rho_\phi$. Knowing the value 
of $\Omega_{GW}$ at $f_0^{BH}$, one then needs to just know the behavior $\Omega_{GW}=f(f_0)$
in the  generic case of a background dominated by a field with an equation of state $w_\phi$.
From 

\beq
\Omega_{GW}(\ahc) \sim \frac{\He^2}{12 \pi^2M_P^2}\frac{k_{\rm hc}^2}{H^2(\ahc)}\sim
\frac{\He^2}{12 \pi^2 M_P^2}\,,
\eeq
we deduce 

\bea
&&
\Omega_{GW}(\abh)=\frac{\He^2}{12 \pi^2 M_P^2}\frac{(\ahc/\abh)^4}{(\ahc/\abh)^{3(1+w_\phi)}}
\nonumber
\\
&&
=\frac{\He^2}{12 \pi^2 M_P^2}\left(\frac{\rhobh}{3 M_P^2}\right)^\frac{1-3w_\phi}{1+3w_\phi}
k_{\rm BH}^\frac{6 w_\phi-2}{3w_\phi+1}
\label{Eq:A3bis}
\,,
\eea

\noi
where we used the fact that $H$ redshifts as $H(a)\propto a^{-3(1+w_\phi)}$, 
$\rho_{GW}\propto a^{-4}$ after horizon crossing, and

\bea
&&
\left(\frac{\abh}{\ahc}\right)^{3(1+w_\phi)}=
\frac{3 M_P^2 H^2(\ahc)}{\rhobh}
=\frac{3 M_P^2 k_{\rm BH}^2(\abh/\ahc)^2}{\rhobh}
\nonumber
\\
&&
\Rightarrow
\frac{\abh}{\ahc}=\left(\frac{3 M_P^2 k^2_{\rm BH}}{\rhobh}\right)^\frac{1}{1+3w_\phi}\,.
\label{Eq:generaleos}
\eea

Note that Eq.~(\ref{Eq:A3bis}) recovers the behavior $\Omega_{GW}=$ cst in the radiation dominated Universe
($w_\phi=\frac13$),
see Eq.~(\ref{Eq:A1}) and $\Omega_{GW}\propto k^{-2}$ for matter dominated ($w_\phi=0$)
case, see Eq.~(\ref{Eq:A2}). For $w_\phi=\frac12$, corresponding to an inflaton potential after inflation $V(\phi)\propto \phi^6$, we obtain $\Omega_{GW}\sim k^\frac25$, which corresponds 
indeed to what we observe in Fig.~\ref{Fig:masterplot}, in the region $A_3$. Combining Eq.~(\ref{Eq:A3bis})
with Eqs.~(\ref{Eq:fbh}) and (\ref{Eq:omegabh}), we obtain for $w_\phi=\frac12$,

\beq
\boxed{
\Omega^{A_3}_{GW}h^2\simeq 10^{-19} \left(\frac{1~\rm g}{\mbh}\right)\left(\frac{10^{-5}}{\beta}\right)^\frac45\left(\frac{f_0}{10^5~\rm Hz}\right)^\frac25\,.
\label{Eq:A3}
}
\eeq

As a conclusion, we analyzed in this section the features of a PGW spectrum 
in a Universe passing successively from an inflaton era, PBH era and radiation dominated era. Ou results are summarized by the Eqs.~(\ref{Eq:A1}), 
(\ref{Eq:A2}) and (\ref{Eq:A3}) in the semiclassical approximation. 
As we mentioned earlier, 
on needs to keep in mind that, despite the challenges of observations \cite{TitoDAgnolo:2024uku,Domcke:2024eti,Aggarwal:2025noe}, a measure of PGW spectrum is a direct and clear insight on the history of pre-BBN physics.

\subsection{Burdening the spectrum}

The memory burden effect affects the shape of the PGW spectrum with several respects, as one can see in Fig.~\ref{Fig:masterplot}. Firstly, the lifetime of the PBHs being longer, the reheating occurs at lower frequencies (smaller Hubble rate). That pushes the transition $\Omega
_{GW}h^2\propto f_0^{-2}\rightarrow \Omega
_{GW}h^2 \sim 10^{-18}$ from $f_{0}^{RH}$
in Eq.~(\ref{Eq:frhnoburden}) to

\beq
\boxed{
\left. f_0^{RH}\right|_n^q=\frac{\alpha^\frac14\sqrt{\epsilon}}{3^\frac34}\frac{2^{\frac{n+1}{2}}}{2 \pi}
\sqrt{3+2n}\left(\frac{M_P}{q \mbh}\right)^{\frac32+n}\left(\frac{g_0}{g_{RH}}\right)^\frac13 T_0\,,
\label{Eq:f0rhburden}
}
\eeq

\noi
where we used Eq.~(\ref{Eq:trhburden}) for $\Trh$. 
Note that we recover the expression (\ref{Eq:frhnoburden}) for $n=0$ and $q=1$.
In our benchmark case ($n=1$, $q=\frac12$), this gives

\bea
&&
\left.f_0^{\rm RH}\right|_{n=1}^{q=\frac12}=
\frac{2^\frac{7}{2}}{3^\frac34}
\frac{\sqrt{5 \epsilon}}{2 \pi} \alpha^\frac14
\left(\frac{g_0}{g_{RH}}\right)^\frac13
\left(\frac{M_P}{\mbh}\right)^\frac52
T_0
\nonumber
\\
&&
\simeq
1.1\times  10^{-1} \left(\frac{1~\rm g}{\mbh}\right)^{\frac52}~\rm Hz\,,
\label{Eq:f0rhburdennq}
 \eea
which is clearly what we see in Fig.~\ref{Fig:masterplot}.

The memory burden also affects the present frequency at which the PBHs begin to dominate the Universe. Indeed, while the formation time remains the same, 
the longer matter
domination era induced by the burdening affects the redshift of the corresponding PGW frequencies. Combining Eq.~(\ref{Eq:f0bhnoburden}) with Eq.~(\ref{Eq:trhburden}),
we obtain

\begin{empheq}[box=\fbox]{align}
\left.f_0^{\rm BH}\right|_n^q=
&
\frac{\alpha^\frac14(3+2n)^\frac16}{3^\frac{5}{12}\pi^\frac23}
\frac{
2^{\frac{n-1}{6}}
\epsilon^\frac16
\beta^{\frac{1+w_\phi}{6 w_\phi}}
}
{
q^{\frac{1}{2}+\frac{n}{3}}
}
\nonumber
\\
&
\times
\left(\frac{M_P}{\mbh} \right)^{\frac56+\frac{n}{3}}
\left(\frac{g_0}{g_{RH}} \right)^\frac13
T_0\,,
\end{empheq}

\noi
which gives, for $q=\frac12$ and $n=1$

\beq
\left. f_0^{\rm BH}\right|_{n=1}^{q=\frac12}\simeq 14\left(\frac{1~\rm g}{\mbh}\right)^{\frac{7}{6}}\sqrt{\frac{\beta}{10^{-8}}}~\rm Hz
\,,
\eeq
which is also what we see in Fig.~\ref{Fig:masterplot}, 
where we took $\beta=10^{-8}$.

\noi
The behavior of $\Omega_{GW} h^2$ between $f_0^{\rm BH}$ and $f_0^{RH}$
is the same with or without the burden case, as in both cases, the Universe is dominated by PBH and is affected by the same redshift,
which is also what we see in Fig.~\ref{Fig:masterplot}.

As a conclusion, we see that the memory burden affects the PGW, acting as a heavier PBH concerning the reheating temperature (because increasing the PBH lifetime), and 
thus pushing $f_0^{\rm RH}$ toward lower values (compare Eq.~(\ref{Eq:f0rhburdennq}) with Eq.~(\ref{Eq:frhnoburden})), but still formed at early
time because the burdening {\it does not} affect the formation and thus the domination time, $\abh$. As a consequence, the memory burden spectrum of PGW is enlarged towards 
{\it lower} frequencies. The gap between $f_0^{\rm RH}$ and $f_0^{\rm BH}$ could then be a clear signature of the presence of a burdening effect in the early Universe.

\section{Inflaton scattering}

\subsection{Generalities}

\begin{figure}[H]
    \centering
    \includegraphics[scale=0.8]{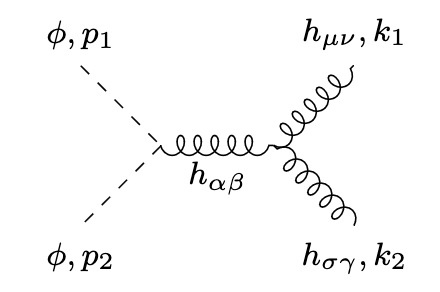}
     \caption{Example of a gravitational process contributing to the gravitational wave background through inflaton scattering.}
    \label{Fig:Feynmaninflaton}
\end{figure}

Another source of graviton involves the inflaton scattering through the exchange of a graviton, see for instance Fig.~\ref{Fig:Feynmaninflaton}
\cite{Choi:2024ilx}. This production of graviton can be interpreted as the high frequency counterpart of fluctuations, that never left the horizon during 
inflation \cite{Ema:2020ggo}. It was shown recently in \cite{Kaneta:2022gug}, and more generically in \cite{Chakraborty:2025zgx} that this Feynman calculation is equivalent to a Bogoliubov approach for modes $k \gg k(\aend)=k_{\rm end}$.

This production mode depends strongly on the equation of state before reheating, through the dilution of the modes $\omega_\phi(a)$ of the inflaton field. Indeed, whereas 
for a quadratic potential, $\omega_\phi={\rm cte}=m_\phi$, for a quartic potential, $\omega_\phi(a) \sim \omega_{\rm end}\left(\frac{\aend}{a}\right)$. 
For a potential $V(\phi) \propto \phi^k$ during reheating, $\omega_\phi(a)\propto \partial^2_\phi V(\phi) \propto a^{-3\frac{k-2}{k+2}}$ \cite{Garcia:2020eof,Garcia:2020wiy,Garcia:2021iag}. It is important to note that 
the process pictured in Fig.~\ref{Fig:Feynmaninflaton} is not, stricto-sensu, a {\it scattering} but a transfer of energy from the oscillations of a {\it classical} background field $\phi$ toward quantum fields, under the form of gravitons in  our case. For a generic potential $\phi^k$, the exact computation necessitates a sum 
over a number of modes $\omega_\phi^n(a), n=1... \infty$ \cite{Clery:2021bwz,Clery:2022wib,Barman:2022qgt,Clery:2023mjo}. The calculation can be tedious, and we describe the method in more detail in the appendix \ref{App:scattering}. This precise calculation with a sum over the modes is the one we used for our numerical analysis. 

However, it was also shown in \cite{Clery:2022wib} that treating the process as a scattering process gives (up to some numerical factors of the order of unity due to the symmetry factors and sum over the modes) a good approximation. It was also the method used in \cite{Ema:2020ggo}. This is the approach we will use in this section to help the reader to understand the shape of the spectrum, the exact result being left in the appendix \ref{App:scattering}.

\subsection{The Boltzmann equation}

To compute the energy spectrum generated by the scattering of the inflaton, on needs to solve the Boltzmann equation

\beq
\frac{d \rho_{GW}}{dt}+4 H \rho_{GW} =2\times n_\phi^2\times \sigma v \times\omega_\phi=\frac{\rho_\phi^2}{4 \pi M_P^4}\omega_\phi\,,
\label{Eq:boltzmann}
\eeq

\noi
where the factor 2 corresponds to 2 gravitons produced per scattering, 
and we used for the amplitude of the process\footnote{In this calculation, we neglected the symmetry factors to simplify the reasoning. For the interested reader, the exact calculation is given in appendix \ref{App:scattering}.}

\beq
|{\cal M}_{\phi \phi \rightarrow hh}|^2 \sim \frac{\omega_\phi^4}{M_P^4}~\Rightarrow ~~\sigma v \sim \frac{|{\cal M}_{\phi \phi \rightarrow hh}|^2}{8 \pi \omega_\phi^2}\sim \frac{\omega_\phi^2}{8 \pi M_P^4}\,,
\nonumber
\eeq
and

\beq
n_\phi^2 \times \sigma v \sim \left(\frac{\rho_\phi}{\omega_\phi}\right)^2 \times \frac{\omega_\phi^2}{8 \pi M_P^4} \sim \frac{\rho_\phi^2}{8 \pi M_P^4}\,.
\eeq

To solve Eq.~(\ref{Eq:boltzmann}), we use the classical change of variable 
$\frac{d}{dt}\rightarrow Ha\frac{d}{da}$, which gives

\beq
\frac{d a^4\rho_{GW}}{da}=\frac{a^3 \rho_\phi^2 ~\omega_\phi(a)}{4 \pi H(a)M_P^4} \,.
\label{Eq:boltzmann(a)}
\eeq

\noi
Instead of integrating Eq.~(\ref{Eq:boltzmann(a)}), we can use it to extract the  spectrum of graviton as function of the present frequency $f_0$.

\subsection{The spectrum (C)}

To obtain the relative spectrum as function of $f_0$, we use

\bea
&&
f_0=\frac{\omega_\phi(a)}{2 \pi}\frac{a}{a_0}=\frac{\omegae}{2 \pi} \left(\frac{\ae}{a}\right)^{3\frac{k-2}{k+2}}\frac{a}{a_0}
\label{Eq:f0}
\\
&&
\Rightarrow d \ln f_0 = \frac{df_0}{f_0}=\left|\frac{8-2k}{k+2}\right| \frac{da}{a}\,,
\label{Eq:dlnf0}
\eea

\noi
which implies

\beq
\frac{d \rho^0_{GW}}{d \ln f_0}= \frac{1}{a_0^4}\frac{d a_0^4 \rho^0_{GW}}{d \ln f_0}
=\frac{1}{a_0^4}\frac{d a^4\rho_{GW}(a)}{d \ln f_0}
\eeq
or, for a potential $V(\phi)\sim \phi^k$,

\beq
\boxed{
\frac{d \rho^0_{GW}}{d \ln f_0}= \frac{1}{4 \pi}\left|\frac{k+2}{8-2k}\right|
\left(\frac{a}{a_0}\right)^4
\frac{\rho_\phi^2 ~\omega_\phi(a)}{H(a)M_P^4}\,,
\label{Eq:drhodlnf0}
}
\eeq

\noi
where we used the fact that $a^4 \rho_{GW}(a)$ is constant, and Eq.~(\ref{Eq:boltzmann(a)}) combined with Eq.~(\ref{Eq:dlnf0}).

During the domination of the inflaton oscillations, we can write $H(a)=\frac{\sqrt{\rho_\phi}}{\sqrt{3}M_P}$. Using Eq.~(\ref{Eq:f0}) 
we deduce that 

\beq
\frac{\ae}{a}=\left(\frac{2 \pi f_0}{\omegae}\right)^\frac{k+2}{2k-8}
\left(\frac{a_0}{\ae}\right)^\frac{k+2}{2k-8}\,.
\label{Eq:aeova}
\eeq

Implementing the expression (\ref{Eq:aeova}) into (\ref{Eq:drhodlnf0}), with $\rho_\phi(a)=\rhoe \left(\frac{\ae}{a}\right)^\frac{6 k}{k+2}$, we obtain

\beq
\frac{d \rho^0_{GW}}{d \ln f_0}=\left|\frac{k+2}{8-2k}\right|
\frac{9 \He^3 \omegae}{4 \pi}\left(\frac{\ae}{a_0}\right)^{-\frac{9}{k-4}}
\left(\frac{2 \pi f_0}{\omegae}\right)^\frac{4k-7}{k-4}\,.
\eeq

\noi
For a quadratic potential, we obtain $\frac{d \rho^0_{GW}}{d \ln f_0}\propto f_0^{-\frac12}$, whereas it is a monochromatic spectrum for a quartic potential\footnote{In fact a sum of monochroms over all the inflaton modes.}. In the case of our
benchmark point, $k=6$, we have

\beq
\frac{d \rho^0_{GW}}{d \ln f_0} = 9\times (2 \pi)^\frac{15}{2}  \He^3 ~\omegae 
\left(\frac{a_0}{\ae}\right)^\frac{9}{2}\left(\frac{f_0}{\omegae}\right)^\frac{17}{2}\,,
\eeq

\noi
or,

\beq
\boxed{
\Omega_{GW}^C h^2=\frac{h^2}{\rho_c}\frac{d \rho^0_{GW}}{d \ln f_0}\simeq 4.2\times 10^{-18} \left(\frac{f_0}{4 \times 10^7~\rm Hz}\right)^\frac{17}{2}\,,
}
\eeq

\noi
where we took for the present critical density $\rho_c=8 \times 10^{47}h^2~\rm GeV^4$,
$\He =2 \times 10^{12}$ GeV, $\omega_\phi=4 \times 10^{13}$ GeV
and $\frac{a_0}{\ae}=10^{30}$. This is effectively what we observe in Fig.~(\ref{Fig:masterplot}), the highest frequency of the spectrum being given by $f_{\rm end}$ in Eq.~(\ref{Eq:fend}).

\section{Scalar fluctuations}

\subsection{Generalities}

PBHs themselves can be a source of gravitational waves. Once dominating the Universe, 
their random distribution generates density fluctuations with a Poisson distribution
 since the locations of black holes are uncorrelated \cite{Papanikolaou:2020qtd,Papanikolaou_2022}. The density (isocurvature) fluctuations 
of PBH, $\delta \rho_{BH}$, generates then potential (curvature) fluctuations $\Phi$ through the Poisson equation

\beq
\nabla^2 \Phi\sim\frac{\delta \rhobh}{M_P^2}
\label{Eq:poisson}
\,.
\eeq
After developing $\Phi$ in its Fourier modes $\Phi_k$, 
Eq.~(\ref{Eq:poisson}) gives

\beq
\Phi_k\sim -\frac{k_{\rm BH}^2}{ k^2}\frac{\delta \rhobh}{\rhobh}\,,
\eeq
where $k_{\rm BH}=H(\abh)$ is the horizon size at the onset of PBHs domination.
A more precise solution of the
Einstein's equation for the gravitational potential, taking into account long wavelength modes, gives \cite{Papanikolaou:2020qtd} 

\beq
\Phi_k\simeq -\left(5+\frac49\frac{k^2}{k_{\rm BH}^2}\right)^{-1}\frac{\delta \rho_{\rm BH}}{\rho_{\rm BH}}\,.
\eeq
Then, the scalar fluctuations $\Phi$  source tensor peturbations $h_{i j}$ of the metric

\beq
ds^2=(1-2\Phi)dt^2-a^2(t)\left[\delta_{ij}(1+2\Phi)+\frac{2~ h_{ij}}{M_P}\right]dx^i dx^j\,,
\eeq
under the form of gravitational waves which follows the equation of motion

\beq
    h''_{k,\lambda} + 2\mathcal{H} h'_{k,\lambda} + k^2 h_{k,\lambda} = \mathcal{S}_{k,\lambda}\,,
    \label{Eq:hij}
\eeq
where the derivatives are with respect to the conformal time $\tau$, $dt=a(\tau)d\tau$ and the source function $\mathcal{S}_{k,\lambda}$ is given in appendix \ref{Sec:densityfluctuation}, Eq.~(\ref{Eq:sklambda}).
For the interested reader, we also describe  in the appendix how we obtain the power spectra from the equation of motion (\ref{Eq:hij}). We discuss in particular the issue of the cutoff scales. In the core of this paper, we want to describe the main features 
of the spectrum, using if necessary rough approximations, such that the reader can understand the behavior of the GW generated by density perturbations, wich corresponds
to the regions $B_i$ of Fig.~\ref{Fig:masterplot}.

\subsection{The spectrum ($B_1$ and $B_2$)}

While we explain in more details how we obtained the spectrum $B_1$ 
in Appendix \ref{Sec:densityfluctuation}, we want in this section 
to help the reader to understand its main shape.
Supposing an almost instantaneous PBH decay, 
it is easy to understand that the higher frequency, $f_{UV}$ 
corresponds roughly to the mean inverse distance $\dbh$ between two PBHs, or 

\bea
&&
f_{UV}(\aev)\sim \frac{1}{2\pi \dbh} \frac{\ain}{\aev}=\frac{1}{2\pi}\left(\frac{4 \pi}{3}\nbh(\ain)\right)^\frac13 \frac{\ain}{\aev}
\nonumber
\\
&&
=\frac{1}{2\pi}\left(\frac{4 \pi}{3}\nbh(\aev)\right)^\frac13
=\frac{2^\frac13}{(3\pi)^\frac23}\left(
\frac{M_P^2\gammabh^2}{\mbh}
\right)^\frac13
\,,
\label{Eq:fuvnoburden}
\eea

\noi
where we used $\nbh(\aev)=\frac{\rhorh}{\mbh}$, with

\beq
\rhorh=3 M_P^2 H^2(\aev)=\frac43M_P^2\gammabh^2\,. 
\eeq
After redshifting $f_{UV}$ from $\arh$ ($\equiv \aev$) to $a_0$, we obtain in, the semiclassical approximation, with $\gammabh$ given by Eq.~(\ref{Eq:tev}), 
and $\trh$ by Eq.~(\ref{Eq:trh}) :

\beq
\boxed{
f_0^{UV}=\left(\frac{g_0}{g_{RH}\pi^2}\right)^\frac13
\left(\frac{\alpha}{3}\right)^\frac14
\left(\frac{\epsilon}{2}\right)^\frac16
\left(\frac{M_P}{\mbh}\right)^\frac56 T_0}
\label{Eq:f0uvnoburden}
\eeq
or
\beq
f_0^{UV}\simeq4.0\times 10^6\left(\frac{1~\rm g}{\mbh}\right)^\frac56~\rm Hz
\,,
\label{Eq:f0uvbis}
\eeq

\noi
which is slightly higher than what we see in Fig.~\ref{Fig:masterplot}. 
This is due to the non linear cutoff we have used in 
order to cut the spectra in our numerical analysis to obtain Fig.~\ref{Fig:masterplot}. 
For the interested reader this issue is discussed in detail 
in the last part of appendix~\ref{Sec:densityfluctuation}. The comparison should be 
made with the most shaded curve of Fig.~\ref{fig:master_plot_cutoffs} which reproduces 
Fig.~\ref{Fig:masterplot} with different cutoffs.
And this fits well with our analytical result (\ref{Eq:f0uvbis}).
The question on how to estimate 
the cutoff is still subject to discussions in the literature \cite{Assadullahi_2009,Inomata:2020lmk,Balaji_2024,Domenech:2024wao}.
We have chosen to 
display the most conservative option in our Fig.~\ref{Fig:masterplot},
but the interested reader can take 
a look at appendix~\ref{Sec:densityfluctuation} and 
Fig.~\ref{fig:master_plot_cutoffs} to appreciate the impact of this choice.


The infrared cutoff of the density fluctuations waves correspond obviously
to the (casually connected) size of the horizon at the time of PBH evaporation, or

\beq
f_{IR}\sim\frac{H(\aev)}{2 \pi}=\frac{1}{3 \pi}\Gammaev\,,
\eeq
which gives in the semiclassical limit, after redshifting from $\arh=\aev$
to $a_0$ :

\beq
\boxed{
f_0^{IR}=\frac{\sqrt{\epsilon} \alpha^\frac14}{3^\frac14 \sqrt{2} \pi}
\left(\frac{M_P}{\mbh}\right)^\frac32
\left(\frac{g_0}{g_{RH}}\right)^\frac13~T_0\,,
}
\eeq
or

\beq
f_0^{IR}\simeq 947~\left(\frac{1~\rm g}{\mbh}\right)^\frac32~\rm Hz
\,,
\eeq
which is also what we clearly see in Fig.~\ref{Fig:masterplot}.

The precise shape of the spectrum between $f_0^{IR}$ and $f_0^{UV}$ is detailed
in appendix \ref{Sec:densityfluctuation}, and necessitates the calculation
of complex second order transfer functions from density perturbations to
scalar (potential) perturbations and finally tensor (metric) 
perturbations. For a complete analysis, we refer the reader to the 
works of \cite{Papanikolaou:2020qtd}, \cite{Domenech:2020ssp,Domenech_2021,Domenech:2024wao} and \cite{Inomata_2019,Bhaumik_2024}. We obtained 

\beq
\Omega^{B_1}_{GW}h^2\propto \left(\frac{f}{f_{UV}}\right)^\frac{11}{3}\,,
\eeq
for $f\lesssim f_{UV}$, and

\beq
\Omega_{GW}^{B_1}h^2\sim\left(\frac{f}{f_{UV}}\right)\,,
\eeq
for $f\gtrsim f_{IR}$, which is also what we observe in Fig.~\ref{Fig:masterplot}, the change of slope corresponding to the passage
from inflaton domination to PBH domination.
The break in the spectrum is more visible in Fig.~\ref{fig:master_plot_cutoffs}.

\subsection{Burdening the spectrum}

The consequence of a memory burden effect on the GW generated by density fluctuations
are clearly visible in Fig.~{\ref{Fig:masterplot}}. As it was the case for the 
PGW, the UV cutoff part of the spectrum is not {\it directly} affected by the burdening, 
because this corresponds roughly to the distance between two PBHs, which is determined at formation time. However, it is the {\it redshift} of this distance
which is affected between $\abh$ and $\arh$, because the PBH domination lasts longer.
Using Eq.~(\ref{Eq:fuvnoburden}), but with the burdened width $\gammabh$ given
in Eq.~(\ref{Eq:mass2}), we obtain

\bea
\left. f_{UV}\right|_n^q
&&=
\frac{2^\frac13}{(3 \pi)^\frac23}\left(\frac{M_P^2 \gammabh^2}{\mbh}\right)^\frac13
\\
&&=M_P
\left(\frac{2^{n+\frac12}(3+2n)\epsilon}{3 \pi ~q^{3+2n}}\right)^\frac23
\left(\frac{M_P}{\mbh}\right)^\frac{7+4n}{3}
\nonumber
\,,
\eea
which gives after redshifting from $\aev$ to $a_0$

\beq
f_0^{UV}=f_{UV}\left(\frac{g_0}{g_{RH}}\right)^\frac13
\frac{T_0}{\trh}
=\frac{\alpha^\frac14}{3^\frac{5}{12}}
\left(\frac{g_0\sqrt{M_P \gammabh}}{\sqrt{2}g_{RH}\pi^2\mbh}\right)^\frac13T_0\,,
\nonumber
\eeq
or
\beq
\boxed{
\left. f_0^{UV}\right|_n^q=\frac{\alpha^\frac14 \epsilon^\frac16}{3^\frac{5}{12}}\frac{2^{\frac{n-1}{6}}(3+2n)^\frac16}{\pi^\frac23 q^{\frac12+\frac{n}{3}}}
\left(\frac{M_P}{\mbh}\right)^{\frac56+\frac{n}{3}}
\frac{g_0^\frac13}{g_{RH}^\frac13}T_0\,.
\label{Eq:f0uvburden}
}
\eeq

Note that we recover the expression (\ref{Eq:f0uvnoburden}) if one takes $n=0$, $q=1$
in Eq.~(\ref{Eq:f0uvburden}). For our benchmark point $n=1$, $q=\frac12$, 
we obtain 

\beq
\left. f_0^{UV} \right|_{n=1}^{q=\frac12} \simeq 1.4 \times 10^5 \left(\frac{1~\rm g}{\mbh}\right)^\frac76~\rm Hz\,,
\eeq
which is again slightly higher than what we can see in Fig.~\ref{Fig:masterplot} due to the cutoff issue already discussed in the unburden case. Nevertheless, this value matches very well with Fig.~\ref{fig:master_plot_cutoffs}, where the most shaded line released the cutoff constraint.

Concerning the IR cutoff, $f_0^{IR}$, it corresponds to the size of the Universe at the time of PBH decay, which is the causally connected region. As in the non-burdened case, this corresponds to $f_0^{RH}$, or

\beq
\left.f_0^{IR}\right|_n^q=\left.f_0^{RH}\right|_n^q
\,,
\eeq

\noi
where $f_0^{RH}|_n^q$ is given by Eq.~(\ref{Eq:f0rhburden}). 
This is indeed what we observe in the region $B_2$ of Fig.~\ref{Fig:masterplot}. In conclusion, the memory burden effect on the GW generated by the density fluctuations of PBHs is also to enlarge the spectrum toward lower frequencies $f_0$. The relation between $f_0^{IR}$
and $f_0^{UV}$ 

\beq
\left.\frac{f_0^{UV}}{f_0^{IR}}\right|_n^q\sim \left(\frac{\mbh}{M_P}\right)^{\frac23(1+n)}q^{1+\frac{2n}{3}}
\,,
\eeq
gives a direct access to the memory burden parameters, allowing to signal the 
presence of such effects on PBHs from the measurement of the corresponding 
frequencies. For instance, the ratio $f_0^{UV}/f_0^{IR}\simeq 4200$ in the semiclassical approach, whereas it
is $\simeq 3.3 \times 10^6$ for our benchmark point $n=1$, $q=0.5$, illustrating the larger

\section{Primordial black holes evaporation}
\label{Sec:PBH evap}
\subsection{The spectrum ($D_1$)}

Another source of gravitational waves is of course the gravitons directly produced by 
the evaporation of the PBHs. If we suppose that the vast majority of them are emitted
at the evaporation time $\gammabh^{-1}$, we expect a peak at the present epoch for a frequency $f_0^{ev}$ corresponding to $\sim \Tbh$ redshifted from $\aev=\arh$ till $a_0$.
The GW energy density spectrum generated by PBHs decay reaches its maximum for the maximum of the distribution function function $g(X)=\frac{X^3}{e^X-1}$, 
which is obtained for $X^{\rm peak}\simeq 2.8$ \cite{Gross:2024wkl}.
Applied to a thermal distribution of gravitons around the PBH,
$d \rho_{h}\sim \frac{k}{2 \pi^2}\times \frac{k^2dk}{e^{\frac{k}{\Tbh}}- 1}$,
this gives

\bea
f_{ev}\simeq\frac{2.8}{2\pi}\Tbh
&&=\frac{ 2.8 M_P^2}{2 \pi \mbh}
\simeq 4.5\times 10^{12}\left(\frac{1 \rm g}{\mbh}\right)~{\rm GeV}
\nonumber
\\
&&
=6.9\times 10^{36}\left(\frac{1~\rm g}{\mbh}\right)~{\rm Hz}
\,,
\label{Eq:fev1}
\eea
which implies 

\beq
f_0^{ev}=f_{ev}\frac{g_0^\frac13}{g_{RH}^\frac13}\frac{T_0}{\trh}
=\frac{(3\alpha)^\frac14}{2^\frac32\pi}\frac{M_P^\frac32}{\mbh \sqrt{\gammabh}}
\frac{g_0^\frac13}{g_{RH}^\frac13}T_0
\,,
\label{Eq:f0evnoburden}
\eeq
where we used Eq.~(\ref{Eq:rhorhpbh}). In the semiclassical limit, where $\gammabh$
is given by Eq.~(\ref{Eq:tev}), we obtain

\begin{empheq}[box=\fbox]{align}
f_0^{ev}&=\left(\frac{\alpha}{3}\right)^\frac14\frac{2.8}{2 \sqrt{2 \epsilon}\pi}\sqrt{\frac{\mbh}{M_P}}\frac{g_0^\frac13}{g_{RH}^\frac13}T_0
\label{Eq:f0ev}
\\
&
\simeq 1.5\times 10^{13}\sqrt{\frac{\mbh}{1~\rm g}}~\rm Hz\,,
\nonumber
\end{empheq}
which is effectively what we observe in the region $D_1$ of Fig.~\ref{Fig:masterplot}.
 
 Note that the evaporation of a black hole does-not produce an exact black body 
spectrum with a fix temperature $\Tbh$. Other factors (called "grey-body" factors) 
need to be taken into account, notably for the possibility of particles reabsorption \cite{Das:2025vts} \footnote{Note that accounting for general relativistic corrections 
to the accretion of relativistic matter onto PBHs can enhance 
their mass growth by an order of magnitude~\cite{Das:2025vts,Maity:2025ffa,Carr:2010wk,Babichev:2005py} shortly after formation, 
leading to important phenomenological implications; throughout this analysis, we 
ignore such effects.}. The temperature $\Tbh$ itself is not a fixed quantity, 
but depends on the proper evolution of $\mbh$ and the redshift 
(and thus on the main component of the Universe). All these effects have of
course been taken into account to produce Fig.~\ref{Fig:masterplot}, and are detailed 
in the appendix \ref{Sec:pbhdecay}. That explains the kind of "misshapen" spectrum
$D_1$, and not a black-body spectrum, we should observe today.

Note also that the gravitational wave amplitude $\rho_{GW}$ at the peak of the spectrum $D_1$,
has a remarkable property: it {\it does not depend} on the mass of the PBH. 
We explain this peculiarity of the spectrum in appendix \ref{Sec:pbhspectrum}, but
also in Fig.~\ref{Fig:differentmasses}, where we considered 
a set of different values for $\mbh$.
Without going into the detail of the spectrum of the products from PBH evaporation, 
we can estimate this peak value, $\Omega_{GW}^{D_1}$, by simple considerations.
Indeed, the total number of particles $N$ produced at $\arh$, should be

\beq
N\sim\frac{\mbh}{\Tbh}= \frac{\mbh^2}{M_P^2}\simeq 5.5\times 10^{10}\left(\frac{\mbh}{1~\rm g}\right)^2\,.
\eeq
From these particles, only $\frac{2}{2+g_{RH}} \times N$ are gravitons, the rest 
populate the thermal bath. 
To estimate the value of $\frac{d\rho_{GW}(f)}{df}$ at 
$f=f_{ev}\simeq \frac{2.8}{2 \pi}\Tbh$, one can then naively expect, 
for a PBH reheating, using Eq.~(\ref{Eq:omegah2}) with $\frac{\rho_{GW}}{\rho_R}=\frac{2}{2+g_{RH}}$

\beq
\boxed{
\Omega_{GW}^{D_1}h^2\simeq 7.5\times 10^{-7}\,,
\label{Eq:D1}
}
\eeq
with $g_{RH}=106.75$.
$\Omega_{GW}^{D_1}$ is of course independent on $\mbh$ or $\beta$, because the PBH 
decay does not distinguish gravitons from radiation, up to the number of degrees of freedom.
The precise calculation and the corresponding peak values are given in appendix \ref{Sec:pbhdecay}.
This value at the peak, as well as
the PGW from inflation Eq.~(\ref{Eq:A1}), is a clear, model independent signature
of a PBH reheating phase. The value corresponds roughly to the one we see in Fig.~\ref{Fig:masterplot}, with some corrections or order unity developed in the appendix \ref{Sec:pbhspectrum}. The main difference comes from the fact that the evaporation is not instantaneous, and $\Tbh$ increases when approaching $\aev$,
diluting the peak value within a range of higher frequencies, as we can see on the left part of point $D_1$ in Fig.~\ref{Fig:masterplot}. 

The limit of this argument comes from the validity of considering $M_{\rm{BH}}$ as 
constant. While this approximation stays valid during most of the PBH evaporation,
this is not the case that close to evaporation. 
All the spectra on the right side of the 
evaporation peak is generated on a timescale much shorter than the Hubble rate. 
During this 
period the rapid increase of the PBH temperature distributes the particles over a much 
larger range of energies, which results in a lower abundance. However since the peak 
is near the point where $\Gamma_{\rm{BH}}^{(n)}\sim H$ we can still get a fair 
estimation of the behavior at the peak, Eq.~(\ref{Eq:D1}), with such a simple argument. 


\subsection{Burdening the spectrum ($D_1'$ and $D_2'$)}

The memory burden effect should of course also affect the GW spectrum from the PBH 
decay. As we discussed in section \ref{Sec:burden}, the memory burden acts as if the evaporation happens in two stages, with a transition when $\mbh=q \Min$.
As a consequence, we expect a spectrum which is a combination of two thermal baths. 
The first phase corresponds to a PBH decaying with a temperature $\Tbh =\frac{M_P^2}{\mbh}$ at a time $\gammabh^{-1}\sim \frac{\Min^3}{3 \epsilon M_P^4}$, 
Eq.~(\ref{Eq:tev}). This frequency peaks at $f_{ev1} \simeq \frac{2.8}{2 \pi}\Tbh$,
with $\Tbh=\frac{M_P^2}{\Min}$, see Eq.~(\ref{Eq:fev1}). However, even if this frequency does not feel directly the memory burden effect, the redshift following the first phase does. In other words, 

\beq
\left. f_0^{ev1}\right|_n^q=\frac{2.8}{2 \pi}\frac{M_P^2}{\Min}
\left(\frac{a_{ev1}}{\arh}\right)\left(\frac{\arh}{a_0}\right)
\nonumber
\eeq
or
\beq
\boxed{
\left. f_0^{ev1}\right|_n^q=\frac{2.8}{2\pi}\frac{\alpha^\frac14}{\sqrt{2\epsilon}}\frac{2^\frac{n}{6}(3+2n)^\frac16}{3^\frac{5}{12}~ q^{\frac12+\frac{n}{3}}}
\left(\frac{M_P}{\Min} \right)^{\frac{n}{3}-\frac12}\frac{g_0^\frac13}{g_{RH}^ \frac13}T_0
\label{Eq:f0ev1}
\,,}
\eeq

\noi
where we used

\beq
\frac{a_{ev1}}{\arh}=\left(\frac{\rhorh}{\rho(a_{ev1})}\right)^\frac{1}{3}=\left(\frac{\gammabh^n}{\gammabh}\right)^\frac23
\,,
\eeq
and

\beq
\frac{\arh}{a_0}=\frac{g_0^\frac13}{g_{RH}^\frac13}\frac{T_0}{\trh}=
\frac{g_0^\frac13}{g_{RH}^\frac13}
\frac{(3 \alpha)^\frac14}{\sqrt{2}}
\frac{T_0}{\sqrt{M_P \gammabh^n}}
\label{Eq:a0ovarh}
\,,
\eeq
$\gammabh^n$ being given by Eq.~(\ref{Eq:mass2}) and $\gammabh$ by Eq.~(\ref{Eq:tev}). Of course, we recover Eq.~(\ref{Eq:f0ev}) for $n=0$, $q=1$, whereas for our benchmark point $n=1$, $q=\frac12$, we obtain 

\beq
\left. f_0^{ev1}\right|_{n=1}^{q=\frac12}\simeq 5.4\times 10^{11}\left(\frac{\mbh}{1 \rm g}\right)^\frac16~\rm Hz\,,
\eeq
which is effectively the first peak $D'_1$ we observe in Fig.~\ref{Fig:masterplot}.

The second peak frequency corresponds to the evaporation of a PBH with a mass $\mbh=q\Min$, redshifted from $\arh$ to $a_0$, or

\beq
f_0^{ev2}=\frac{2.8}{2 \pi}\frac{M_P^2}{q\Min}\frac{\arh}{a_0}
\eeq
which gives, using (\ref{Eq:a0ovarh}),

\beq
\boxed{
\left.f_0^{ev2}\right|_n^q=\frac{2.8}{2 \pi}\frac{(3 \alpha)^\frac14}{\sqrt{2 \epsilon}}\frac{1}{2^{\frac{n}{2}}\sqrt{3+2n}}\left(\frac{M_P}{q\Min}\right)^{-\frac12-n}\frac{g_0^\frac13}{g_{RH}^\frac13}T_0\,,
\label{Eq:f0ev2}
}
\eeq
where we recover also Eq.~(\ref{Eq:f0ev}) for $n=0$, $q=1$, whereas for our benchmark point $n=1$, $q=\frac12$, we obtain 

\beq
\left.f_0^{ev2}\right|_{n=1}^{q=\frac12}\simeq 7\times 10^{17}\left(\frac{\mbh}{1~\rm g}\right)^\frac32~\rm Hz\,,
\eeq
which corresponds to the second peak, $D'_2$ we see in Fig.~\ref{Fig:masterplot}. 

Note that increasing $\mbh$ pushes the frequency of the first and second peaks toward larger values, 
but with a different dependence on $\mbh$. In another words, a measurement of the two peaks is not only a proof of the existence of a memory burden effect, but also
can help to determine the parameters $n$ and $q$. Indeed, if a double--peaked
signal is observed, the ratio between the two frequencies 

\beq
\frac{f^{ev2}_0}{f^{ev1}_0}\simeq \frac{1.2 q^{1+\frac43}n}{(2^n(3+2n))^\frac23}
\left(\frac{\Min}{M_P}\right)^{\frac43n-1}\sim \left(\frac{\Min}{M_P}\right)^{\frac43n-1}
\,,
\eeq
combined with Eq.~(\ref{Eq:f0ev2}), gives a direct access to $n$.

Concerning the amplitude, the second phase gives rise to the reheating process, and then, all the arguments used to compute $\Omega_{GW}^{D_1}$ are also valid here.
A more detailed calculation gives a small dependence of the peak value on the burden parameter $n$, and is presented in appendix \ref{Sec:pbhspectrum}.
However, the unburdened computation stay a good approximation even in this case, 
or

\beq
\boxed{
\Omega_{GW}^{D'_2}=\Omega_{GW}^{D_1}\simeq 7.5\times 10^{-7}\,.
}
\eeq

\noi
This is not the case for the lower frequency peak, generated by the first phase of the evaporation, due to the redshift of $\rho_{GW}$ in a matter dominated Universe before $\arh$.

\section{Dependance on $\mbh$ and $w_\phi$} 

\begin{figure}
    \centering
    \includegraphics[width=\linewidth]{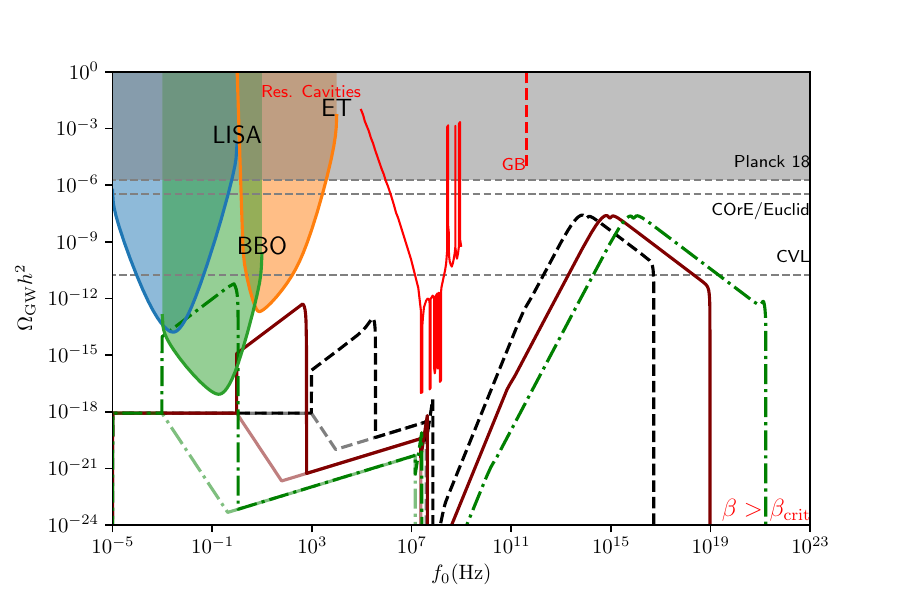}
    \caption{GW spectra in the unburden case with $w_{\phi}=1/2$ for $M = 1,10^{2}\rm{\,and \,} 10^{4}g$ respectivelly for black, red and green with $\beta = 10^{-5},10^{-6}\rm{\,and \,}10^{-7}$}
    \label{Fig:differentmasses}
\end{figure}

\begin{figure}
    \centering
    \includegraphics[width=\linewidth]{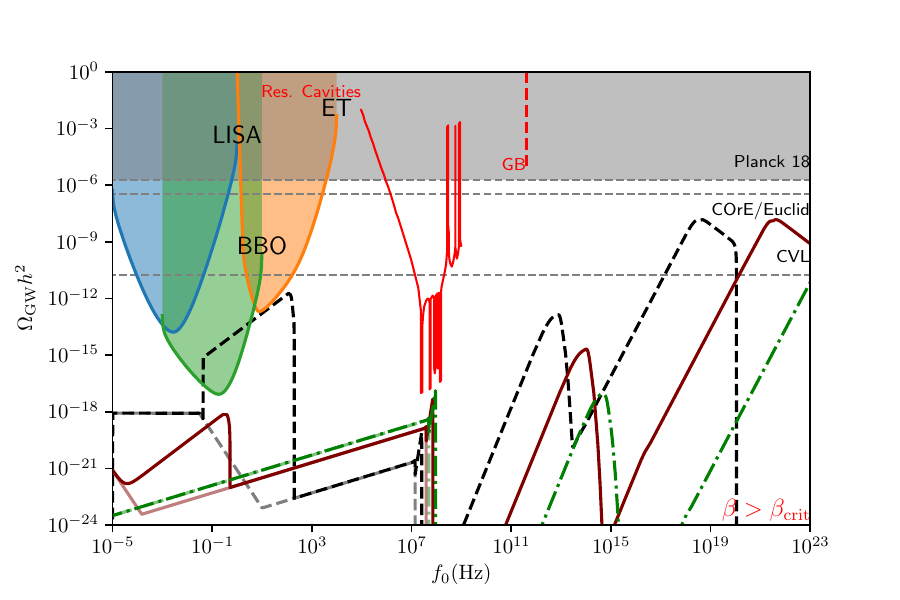}
    \caption{GW spectra in the burden case with $w_{\phi}=1/2$ for $M = 1,10^{2}\rm{\,and \,} 10^{4}g$ respectivelly for black, red and green with $\beta = 10^{-8},10^{-13}\rm{\,and \,}10^{-17}$}
    \label{Fig:differentmassesburden}
\end{figure}

Our study focused on the benchmark point $\mbh=1$ g and $w_\phi=\frac12$, 
to help the reader understand the different regions of the PGW spectrum represented in Fig.~\ref{Fig:masterplot}. We want in this section to complete 
our study to larger masses
and different values of $\beta$. 
We show our result for $\mbh=100$ and $10^4$ g in Fig.~\ref{Fig:differentmasses},
(Fig.~\ref{Fig:differentmassesburden}) in the unburden (burden) case, 
from the black-dashed line to green-dash-dotted line. 

As already discussed 
in the previous section, concerning the PBH evaporation point $D_1$,
we observe a slight shift toward larger frequencies for larger masses.
This comes from the fact that heavier PBH decays later, limiting the 
redshift of the gravitons produced from $\arh$ to $a_0$. On the other hand, for the density fluctuation peak (regions $B_1$ and $B_2$), we see a large shift toward {\it lower} frequencies 
for larger $\mbh$. This is due to the fact that the horizon is larger at evaporation time $\arh$ for heavier PBHs. The distance between them, as well as the size of the horizon is then larger, pushing $f_0^{IR}$ and $f_0^{UV}$ towards {\it lower} values.

Concerning the primordial gravitational waves, increasing the mass of PBHs extend
the inflaton domination region $A_3$ because heavier PBH are formed later. As a consequence, the frequency where the PBH dominates the energy budget of the Universe,
$f_0^{BH}$ is shifted naturally toward lower values, as one can see in the left part of Fig.~\ref{Fig:differentmasses} and Eq.~(\ref{Eq:fbh}).
The inflaton scattering can only be modified through the overall redshift effect induced by the PBH domination era which then follows the behavior of the PGW dependence.

If one takes into account the memory burden effect, the conclusions are largely the same, as we show in Fig.~\ref{Fig:differentmassesburden}. Whereas the two peaks generated by the PBHs decay are shifted towards larger frequencies for larger mass due to a reduction of the redshift, the density perturbation spectrum is shifted toward even much lower frequencies than in the non-burden case due to an effective enlargement of the PBH lifetime.

\begin{figure}
    \centering
    \includegraphics[width=\linewidth]{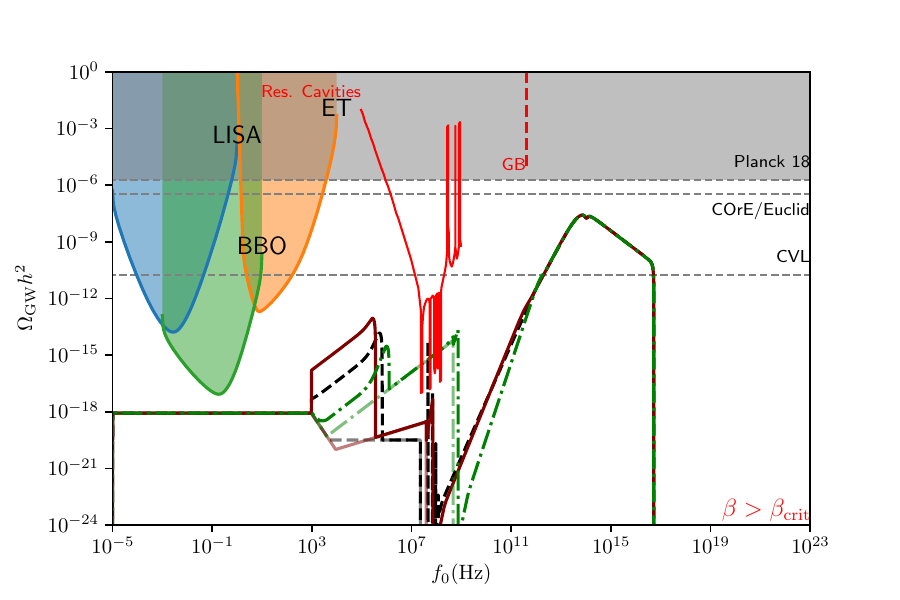}
    \caption{GW spectra in the unburdened case with $w_{\phi}=1/3,1/2\rm{\,and \,}1 $ respectivelly for black, red and green with $\beta = 10^{-4},10^{-5}\rm{\,and \,}10^{-9}$. The PBH mass is fixed to $M = 1g$}
    \label{Fig:beta}
\end{figure}

\begin{figure}
    \centering
    \includegraphics[width=\linewidth]{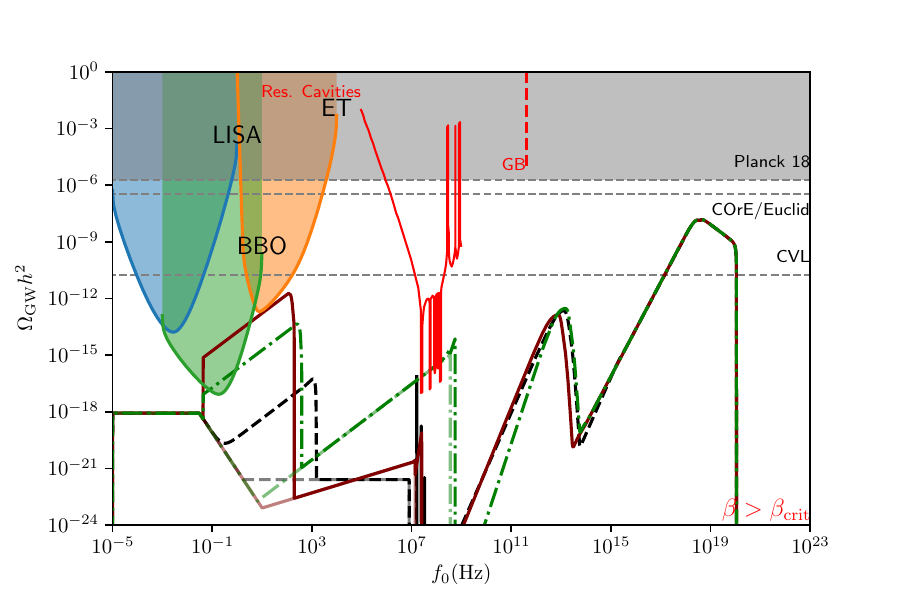}
    \caption{GW spectra in the burden case with $w_{\phi}=1/3,1/2\rm{\,and \,}1 $ respectivelly for black, red and green with $\beta = 10^{-7},10^{-8}\rm{\,and \,}10^{-13}$. The PBH mass is fixed to $M = 1g$}
    \label{Fig:betaburden}
\end{figure}

The effect of the value of $\beta$ is much weaker, as we show in Fig.~\ref{Fig:beta}. Whereas the PBH decay spectrum is mainly unchanged, due to the universal nature of the decay into gravitons once PBHs dominate the Universe, as we explained above, the density perturbation spectrum is altered. Smaller $\beta$ means later domination, or a longer $\phi$ domination time. In Fig.~\ref{Fig:beta}, we also show the influence of the EoS $w_\phi$ on the GW spectrum for a fixed $\mbh=1$ g. The slope of the spectrum range from
$\Omega_{GW}^{A_3}\propto f_0^{-2}$ before the PBH formation for $w_\phi=0$, to $\Omega_{GW}^{A_3}\propto f_0$ for $w_\phi=1$, 
which is in agreement with our result (\ref{Eq:generaleos}), $\Omega_{GW}^{A_3}\propto f_0^{\frac{6 w_\phi-2}{3 w_\phi+1}}$. Taking into account the memory burden effect, we obtained Fig.~\ref{Fig:betaburden} for different values of $w_\phi$.
The dependence on $w_\phi$ is comparable to the non-burden case.

\section{Conclusion}

Our conclusion can be summarized by Fig.~\ref{Fig:masterplot}, where we show
all the sources we took into account to obtain a PGW spectrum in presence of the inflaton field and a population of PBH.
Indeed, by its proper nature, gravitational waves have many sources, because any object couples in a way or another, with the metric. In our work, 
we decided to concentrate on two main sources of gravitational waves : the primordial black holes, and the inflaton field. Our choice was dictated by the fact that these two sources can be considered as a minimal as long as one 
considers the possibility of inflation and the existence of primordial black hole.
Any other extension would generate a spectrum {\it above} the one we presented
in Fig.~\ref{Fig:masterplot}. We also extended our analysis, considering 
the possibility of memory burden effects in the process of PBH evaporation.

We then distinguished typical signatures in the spectrum, characteristic of the 
different gravitational wave sources. Having access to the spectrum of primordial 
gravitational waves is like reading the history of the Universe from inflation to the CMB. We can guess the species which dominates the evolution at each stage, recognize the presence of primordial 
holes, and even measure the reheating temperature. For all these reasons, we consider 
that the hunt for primordial gravitational waves is one of the major objectives of our discipline for the decades to come.

\section*{Acknowledgments}
The authors want to warmly thank Donald Kpatcha, Simon Clery, Vincent Vennin and Lucien Heurtier for extremely useful discussions, especially during the Astro@Paris-Saclay symposium 2024 and the "Gravitational Probes of the Early Universe" workshop at King's College, London in 2025. 
This project has received funding from the European Union’s Horizon Europe research and innovation programme under the Marie Skłodowska-Curie Staff Exchange grant agreement No 101086085 – ASYMMETRY and the CNRS-IRP project UCMN. MRH thanks ISI, Kolkata, for financial support through Research Associateship.

\section*{Appendix}

\appendix

\section{Primordial gravitational wave spectrum: a primer}
\label{App:pgw}

\subsection{The spectrum}

For the reader interested by a detailed analysis of the PGW spectrum, 
we refer to \cite{Boyle:2005se,Watanabe:2006qe,Saikawa:2018rcs,Caprini:2018mtu,Bernal:2019lpc,Haque:2021dha,Maity:2024cpq,Barman:2023ktz}. 
The aim of this appendix is to give to the reader the necessary tools to 
understand section \ref{sec: 3}. Here, primordial gravitational waves refer to those generated as tensor 
perturbations due to vacuum fluctuations during inflation, in 
the absence of any source terms. These fluctuations are described by the perturbed FLRW metric
\bea
\mathrm{d}s^2 = a^2(\eta)\left[\mathrm{d}\eta^2 - (\delta_{ij} + h_{ij})\, \mathrm{d}x^i \mathrm{d}x^j \right]\,,
\eea

\noi
where $h_{ij}$ is transverse and traceless. In Fourier space, the mode equation for $h_k$ reads

\bea
h_k'' + 2\frac{a'}{a} h_k' + k^2 h_k = 0\,.
\eea
Defining the Mukhanov-Sasaki variable $u_k = a h_k$, the evolution becomes
\bea
u_k'' + \left(k^2 - \frac{a''}{a} \right) u_k = 0\,.
\eea
Assuming de Sitter inflation, the scale factor is $a_I(\eta) = -1/(H_{\rm I} \eta)$\footnote{Note that we assume the Hubble parameter remains approximately constant during inflation. For slow-roll inflationary models, its variation is negligible, so we set its value at the end of inflation, denoted as $H_{\rm I} \simeq H_{\rm end}$.}, yielding a nearly scale-invariant power spectrum

\bea
\mathcal{P}_T(k) = \frac{4k^3}{2\pi^2}|h_k|^2 \simeq \frac{H_{\rm I}^2}{2\pi^2 M_{P}^2} \left(1 + \frac{k^2}{k_{\rm end}^2} \right),
\label{Eq:pgwspectrum}
\eea

\noi
where $k_{\rm end}$ is the comoving scale crossing the 
horizon at the end of inflation. We focus on the leading term $k \ll k_{\rm end}$, where the 
spectrum is nearly flat. To track GW evolution post-inflation, we factor out the inflationary mode amplitude: $h_k(\eta) = h_k(\eta_{\rm end})\,\chi_k(\eta)$, 
where $\chi_k$ evolves through reheating and radiation eras. The transfer function satisfies:
\bea
\chi_k'' + 2 \frac{a'}{a} \chi_k' + k^2 \chi_k = 0\,.
\label{Eq:evolutionchi}
\eea
During reheating, dominated by an equation of state $w_\phi$, the Hubble parameter evolves as $H = H_{\rm end} (a/a_{\rm end})^{-\frac{3(1+w_\phi)}{2}}$. The Eq.~(\ref{Eq:evolutionchi}) then becomes

\bea
\chi_k'' + \frac{5 - 3w_\phi}{2\,(a/a_{\rm end})} \chi_k' + \frac{(k/k_{\rm end})^2}{(a/a_{\rm end})^{1-3w_\phi}} \chi_k = 0,
\eea

\noi
whose solution is the Bessel-function

\bea
\chi_k^{\rm RH}(a) = (a/a_{\rm end})^{-\nu} \left[ C_{1k} J_{-\nu/\gamma}(x) + C_{2k} J_{\nu/\gamma}(x) \right]\,,
\nonumber
\eea
where $\nu = \frac{3}{4}(1 - w_\phi)$, $\gamma = \frac{1}{2}(1 + 3w_\phi)$, and $x = k (a/a_{\rm end})^\gamma / (\gamma k_{\rm end})$. Matching conditions at the end of inflation determine $C_{1k}$ and $C_{2k}$,
\begin{align}
C_{1k} &= \frac{\pi^2}{2\gamma k_{\rm end}} \left[ 
\frac{k}{ik-k_{\rm end}} J_{\frac{\nu}{\gamma}}\left(\frac{k}{\gamma k_{\rm end}}\right) 
- J_{\frac{\gamma+\nu}{\gamma}}\left(\frac{k}{\gamma k_{\rm end}}\right) 
\right]\\ \nonumber &\csc\left(\frac{\pi\nu}{\gamma}\right), \\
C_{2k} &= \frac{\pi^2}{2\gamma k_{\rm end}} \left[ 
\frac{k}{ik-k_{\rm end}} J_{-\frac{\nu}{\gamma}}\left(\frac{k}{\gamma k_{\rm end}}\right) 
- J_{-\frac{\gamma+\nu}{\gamma}}\left(\frac{k}{\gamma k_{\rm end}}\right) 
\right]\\ \nonumber &\csc\left(\frac{\pi\nu}{\gamma}\right).
\end{align}
 
 \noi
 where csc stands for the cosecant of the product $\pi\nu/\gamma$, 
$\csc(\pi \nu /\gamma) = \frac{1}{\sin(\pi \nu /\gamma)}$.
During radiation domination, $H \propto a^{-2}$, and the solution becomes
\begin{equation}
\chi_k^{\rm RD}(a) = \frac{1}{\frac{a}{a_{\rm end}}} \left[D_{1k} e^{\frac{-i k \frac{a}{a_{\rm end}} }{k_{\rm RH} \frac{a_{\rm RH}}{a_{\rm end}}}}+ D_{2k} e^{\frac{i k \frac{a}{a_{\rm end}}} {k_{\rm RH} \frac{a_{\rm RH}}{a_{\rm end}}}} \right]\,.
\end{equation}
As usual, the coefficients $D_{1k}$ and $D_{2k}$ are determined by matching the mode solutions and their derivatives at reheating, and are given by
\begin{align}
D_{1k} &= \frac{a_{\rm RH}/a_{\rm end}}{2} e^{i k / k_{\rm RH}} \bigg[ \left(1 + \frac{i k_{\rm RH}}{k}\right) \chi_k^{\mathrm{RH}}(a_{\rm RH}) \\\nonumber
&\qquad + i \frac{k_{\rm RH}}{k} \frac{a_{\rm RH}}{a_{\rm end}} \frac{\partial \chi_k^{\mathrm{RH}}(a_{\rm RH})}{\partial (a/a_{\rm end})} \bigg]
= i \frac{(a_{\rm RH}/a_{\rm end}) k_{\rm RH}}{2 k} \mathcal{E}_{1k}.\\
D_{2k} &= \frac{a_{\rm RH}/a_{\rm end}}{2} e^{-i k / k_{\rm RH}} \bigg[ \left(1 - \frac{i k_{\rm RH}}{k}\right) \chi_k^{\mathrm{RH}}(a_{\rm RH}) \\\nonumber
&\qquad - i \frac{k_{\rm RH}}{k} \frac{a_{\rm RH}}{a_{\rm end}} \frac{\partial \chi_k^{\mathrm{RH}}(a_{\rm RH})}{\partial(a/a_{\rm end}) } \bigg]
= -i \frac{(a_{\rm RH}/a_{
\rm end}) k_{\rm RH}}{2 k} \mathcal{E}_{2k}.
\end{align}
Deep inside the radiation-dominated era, the gravitational wave energy density per logarithmic interval is given by:
\bea
\rho_{GW}(k,\eta) = \frac{M_{\rm p}^2 k^2 a_{\rm end}^2}{4 a^4} \mathcal{P}_T(k) (|D_{1k}|^2 + |D_{2k}|^2).
\eea

\noi
Normalizing to the critical density $\rho_c = 3M_{P}^2 H^2$, the dimensionless density parameter is:
\bea
\Omega_{GW}(k, \eta) = \frac{\mathcal{P}_T(k)}{48} \left(|\mathcal{E}_{1k}|^2 + |\mathcal{E}_{2k}|^2 \right).
\eea
At late times, the redshifting leads to
\bea
\Omega_{ GW}(k) h^2 = c_g\, \Omega_{\rm rad,0} h^2 \,\Omega_{ GW}(k, \eta),
\eea
where $c_g\sim 0.4$ accounts for changes in relativistic degrees of freedom. The current radiation energy density, including photons and three neutrino species, is given by $\Omega_{\rm rad,0} h^2 = 4.16 \times 10^{-5}$. In the long-wavelength limit $k \ll k_{\rm RH}$, the mode functions satisfy $|{\cal E}_{1k}|^2 = |{\cal E}_{2k}|^2 \simeq 1$, indicating no spectral distortion. However, in the short-wavelength limit $k \gg k_{\rm RH}$, the amplitude exhibits scale-dependent enhancement,
\bea
|{\cal E}_{1k}|^2 = |{\cal E}_{2k}|^2 = \frac{4\gamma^2}{\pi} \Gamma^2\left(1+\frac{\nu}{\gamma}\right) \left(\frac{k}{2\gamma k_{\rm RH}}\right)^{n_{GW}},
\eea
where the spectral index is
\bea
n_{ GW} = 1 - \frac{2\nu}{\gamma} = -\frac{2(1 - 3w_\phi)}{1 + 3w_\phi}.
\eea
This spectral tilt arises from the redshifting of modes relative to the background during reheating and is determined by the effective equation of state $w_\phi$.
The present-day GW energy density spectrum is then given by
\bea
\Omega_{GW}^{\rm rad}(k)\,h^2 = \frac{\Omega_{\rm rad,0} h^2}{12\pi^2} \frac{H_{\rm end}^2}{M_{P}^2}, \quad \text{for } k \ll k_{\rm RH},
\eea
and in the high-frequency regime $k \gg k_{\rm RH}$, it becomes
\begin{equation}
\Omega_{ GW}(k)\,h^2 = \Omega_{GW}^{\rm rad}(k)\,h^2 \cdot \frac{4\gamma^2}{\pi} \Gamma^2\left(1+\frac{\nu}{\gamma}\right) \left(\frac{k}{2\gamma k_{\rm RH}}\right)^{n_{ GW}}.
\label{eq: gw energy density final}
\end{equation}
This expression encapsulates the effect of reheating dynamics on the high-frequency tail of the GW spectrum, which 
varies with the background equation of state: it is red-tilted ($n_{GW} < 0$) for $w_\phi < 1/3$, scale-invariant ($n_{ GW} = 0$) for $w_\phi = 1/3$, and blue-tilted ($n_{GW} > 0$) for $w_\phi > 1/3$. For example, matter-like reheating with $w_\phi = 0$ leads to $n_{ GW} = -2$, while a kination-dominated phase with $w_\phi = 1$ yields $n_{GW} = 1$.

\subsection{Constraints on reheating}

For a stiff post-inflationary equation of state, the reheating temperature $T_{\rm RH}$ can be constrained by the additional relativistic degrees of freedom sourced by gravitational waves. We now analyze how small $T_{\rm RH}$ can be while remaining consistent with bounds on $\Delta N_{\rm eff}$, assuming fixed values of the equation of state $w_\phi$ and the inflationary energy scale. Here, $\Delta N_{\rm eff}$ quantifies the excess relativistic degrees of freedom at the epochs of Big Bang Nucleosynthesis (BBN) or CMB decoupling.
For a stochastic GW background, the contribution to $\Delta N_{\rm eff}$ is given by~\cite{Jinno:2012xb}
\begin{equation}
\Delta N_{\rm eff} = \frac{\rho_{ GW}}{\rho_\nu}
= \frac{8}{7} \left( \frac{11}{4} \right)^{4/3} \frac{\rho_{GW}}{\rho_\gamma}\,,
\label{Eq: neff}
\end{equation}
where $\rho_\nu$ and $\rho_\gamma$ denote the energy densities of a single neutrino species and photons, respectively. This expression uses the standard relation $T_\nu = \left(4/11\right)^{1/3} T_\gamma$. Equation~\eqref{Eq: neff} imposes an upper limit on the total GW energy density today
\begin{equation}
\int_{k_{\rm RH}}^{k_{\rm end}} \frac{dk}{k} \, \Omega_{GW}(k) h^2
\leq \frac{7}{8} \left( \frac{4}{11} \right)^{4/3} \Omega_\gamma h^2 \, \Delta N_{\rm eff},
\label{eq: deltaneff}
\end{equation}
where the present-day photon energy density is $\Omega_\gamma h^2 \simeq 2.47 \times 10^{-5}$.
For blue-tilted spectra ($w_\phi > 1/3$), the GW energy is dominated by modes near $k_{\rm end}$, which exited the Hubble horizon at the end of inflation and reentered shortly after. In this regime, substituting the GW spectrum from Eq.~~\eqref{eq: gw energy density final} into Eq.~~\eqref{eq: deltaneff} and integrating yields
\begin{equation}
\int_{k_{\rm RH}}^{k_{\rm end}} \frac{dk}{k} \, \Omega_{ GW}(k) h^2
\simeq \Omega_{GW}^{\rm rad} h^2 \, \zeta(w_\phi) \left( \frac{k_{\rm end}}{k_{\rm RH}} \right)^{\frac{6w_\phi - 2}{1 + 3w_\phi}},
\label{eq: BBNapprox}
\end{equation}
where
\begin{equation}
\zeta(w_\phi) = (1 + 3w_\phi)^{\frac{4}{1 + 3w_\phi}}
\Gamma^2\left( \frac{5 + 3w_\phi}{2 + 6w_\phi} \right)
\frac{1 + 3w_\phi}{2\pi(3w_\phi - 1)}.
\end{equation}
The ratio between the comoving scales $k_{\rm end}$ and $k_{\rm RH}$ is determined by reheating dynamics:
\begin{eqnarray}
\frac{k_{\rm end}}{k_{\rm RH}} =
\left( \frac{30 \, \rho_{\rm end}}{\pi^2 g_{RH}} \right)^{\frac{1 + 3w_\phi}{6(1 + w_\phi)}}
T_{\rm RH}^{-\frac{2}{3} \cdot \frac{1 + 3w_\phi}{1 + w_\phi}},
\label{eq: kend kre}
\end{eqnarray}
where $\rho_{\rm end} = 3 M_{\rm p}^2 H_{\rm end}^2$ is the energy density at the end of inflation. Combining Eqs.~\eqref{eq: deltaneff}, \eqref{eq: BBNapprox}, and \eqref{eq: kend kre}, we obtain a lower bound on the reheating temperature:
\begin{equation}
T_{\rm RH} \geq
\left( \frac{\Omega_{GW}^{\rm rad} h^2 \, \zeta(w_\phi)}{5.61 \times 10^{-6}\, \Delta N_{\rm eff}} \right)^{\frac{3(1 + w_\phi)}{4(3w_\phi - 1)}}
\left( \frac{30 \, \rho_{\rm end}}{\pi^2 g_{RH}} \right)^{1/4}\,.
\label{eq: BBNrestriction}
\end{equation}

\noi
Here, we assume that the degrees of freedom associated with the thermal bath and entropy are the same at the end of reheating, denoted by $g_{\rm RH}$.

\subsection{PBH domination}

Up to this point, we have considered a scenario where PBHs do not dominate the energy 
density before evaporating. However, if the initial PBH fraction $\beta$ exceeds a critical threshold $\beta_{\rm c}$, the constraint on the reheating temperature 
$T_{\rm RH}$ given in Eq.~\eqref{eq: BBNrestriction} can be relaxed. This is because in a PBH-dominated phase, typically a matter-dominated era sandwiched between a 
radiation-dominated (RD) phase and a $w_\phi$ dominated phase, the spectral behavior of the gravitational wave relic changes, with $\Omega_{\rm GW}\,h^2 \sim k^{-2}$. As 
a result, the stringent constraints from $\Delta N_{\rm eff}$, can be avoided if the PBH domination lasts sufficiently long. A rough consistency check with the 
$\Delta N_{\rm eff}$ bound can be performed by evaluating the gravitational wave relic abundance at the high-frequency end of the spectrum, $k_{\rm end}$. The requirement is that it satisfies the inequality:
\begin{equation}\label{eq: omegaineq}
\Omega_{GW}(k_{\rm end})\,h^2 \leq \frac{7}{8} \left( \frac{4}{11} \right)^{4/3} \Omega_\gamma h^2\, \Delta N_{\rm eff} \sim 9.5 \times 10^{-7}\,,
\end{equation}
\noi
where we have used the latest bound on $\Delta N_{\rm eff}$ from the P-ACT-LB dataset, which imposes the constraint $\Delta N_{\rm eff} < 0.17$ at $95\%$ confidence level \cite{ACT:2025fju,ACT:2025tim}. In the presence of PBH domination, the spectral behavior of gravitational waves during the $w_\phi$-dominated era is

\begin{equation}
\Omega_{GW}(k)\,h^2 \sim \Omega_{GW}^{\rm rad}\,h^2 \cdot \frac{{\Gamma\left(\frac{5}{2}\right)}^2}{\pi} \left(\frac{k_{\rm BH}}{k_{\rm RH}}\right)^{-2} \left(\frac{k}{k_{\rm BH}}\right)^{n_{\rm GW}}\,.
\label{eq: pbh-spectrum}
\end{equation}

Therefore, to estimate the gravitational wave relic abundance at $k_{\rm end}$, one simply substitutes $k = k_{\rm end}$ in the above expression. The relevant wavenumber ratios in this context take the following form:

\begin{eqnarray}
&\frac{k_{\rm BH}}{k_{\rm RH}} =\sqrt{2} \beta^{\frac{1 + w_\phi}{6 w_\phi}} \cdot \left( \frac{2\pi\,\gamma \,M_{\rm in}^2}{\epsilon \,M_{P}^2} \right)^{1/3}\,,\\&
\frac{k_{\rm end}}{k_{\rm BH}} = \left( \frac{a_{\rm BH}}{a_{\rm end}} \right)^{\frac{1+3w_\phi}{2}},
\end{eqnarray}
where, $\left( \frac{a_{\rm BH}}{a_{\rm end}} \right) = \left( \frac{M_{\rm in} \sqrt{\rho_{\rm end}}}{4\pi\gamma \sqrt{3} M_{\rm p}^3} \right)^{\frac{2}{3(1 + w_\phi)}}\,\beta^{-\frac{1}{3\,w_\phi}}$. Upon substituting the wave number ratios into the gravitational wave spectrum expression, Eq.\eqref{eq: pbh-spectrum}, one obtains

\begin{eqnarray}
\Omega_{GW}(k_{\rm end})\, h^2 \simeq \Omega_{GW}^{\rm rad}\, h^2 \cdot 
\frac{3^{\frac{7 + 3w_\phi}{3(1 + w_\phi)}}}{2^{\frac{13 + 29w_\phi}{3(1 + w_\phi)}}}
\left( \frac{M_{P}}{M_{\rm in}} \right)^{\frac{2(3 - w_\phi)}{3(1 + w_\phi)}} \\\nonumber 
\times \frac{\epsilon^{2/3}}{(\pi \gamma)^{\frac{8w_\phi}{3(1 + w_\phi)}}}
\left( \frac{\rho_{\rm end}}{M_P^4} \right)^{\frac{3w_\phi - 1}{3(1 + w_\phi)}}
\beta^{-\frac{4}{3}}\,.
\end{eqnarray}

The above relic abundance must remain below the upper limit set by the bound on $\Delta N_{\rm eff}$, which is approximately $10^{-6}$, corresponding to $\Delta N_{\rm eff} = 0.17$. This constraint on the gravitational wave relic abundance can be translated into an upper bound on the reheating temperature. Utilizing Eq.~\eqref{Eq:rhorhpbh}, one finds

\begin{eqnarray}
\frac{T_{\rm RH}}{M_P} \leq 
\frac{2^{\frac{5(9 + 17w_\phi)}{4(3 - w_\phi)}}}{3^{\frac{9 + 5w_\phi}{2(3 - w_\phi)}}}
\left( \frac{10^{-6}}{\Omega_{ GW}^{\rm rad}\, h^2} \right)^{\frac{9(1 + w_\phi)}{4(3 - w_\phi)}}
\left( \frac{\rho_{\rm end}}{M_P^4} \right)^{\frac{3(1 - 3w_\phi)}{4(3 - w_\phi)}}
\\\nonumber \times \beta^{\frac{3(1+w_{\phi})}{3-w_{\phi}}}\left( \frac{\pi \gamma}{\epsilon^{1/3}} \right)^{\frac{6w_\phi}{3 - w_\phi}} 
\alpha^{-1/4}\,.
\end{eqnarray}

\subsection{Including the memory burden effect}

Now, let us modify the scenario by considering a $phase\, of \,memory\, burden$ that sets in after the black hole has lost a fraction $q$ of its initial mass, instead of using the standard semiclassical approximation (i.e., Hawking evaporation). In this framework, the mass dissipation rate is suppressed by the black hole entropy as $\sim1/ S^n$ (for details, see the next section). Under this assumption, the ratio of the relevant wave numbers is modified and can be expressed as:
\begin{align}\label{eq:ratiok}
    \frac{k_{\rm BH}}{k_{\rm RH}}&=\begin{cases}
        \sqrt{2}\beta^{\frac{1+w_\phi}{6w_\phi}}\left[\frac{6\pi\gamma}{2^n\epsilon(3+2n)}\left(\frac{qM_{\rm in}}{M_P}\right)^{2(1+n)}\right]^{\frac{1}{3}},\, {\rm for}\, \beta>\beta_\star,\\[10pt]   \sqrt{2}(q\beta)^{\frac{1+w_\phi}{6w_\phi}}\left[\frac{6\pi\gamma}{2^n\epsilon(3+2n)}\left(\frac{qM_{\rm in}}{M_P}\right)^{2(1+n)}\right]^{\frac{1}{3}}, {\rm for}\, \beta<\beta_\star. 
    \end{cases}
\end{align}
and 
\begin{align}\label{eq:ratiok1}
    \frac{k_{\rm end}}{k_{\rm BH}}&=\begin{cases}
        \frac{\beta^{-\frac{1+3w_\phi}{6w_\phi}}}{\left(4\pi\sqrt{3}\gamma\right)^{\frac{1+3w_\phi}{3(1+w_\phi)}}}\left(\frac{M_P}{M_{\rm in}}\right)^{-\frac{1+3w_\phi}{3(1+w_\phi)}}\left(\frac{\rho_{\rm end}}{M_P^4}\right)^{\frac{1+3w_\phi}{6(1+w_\phi)}},\\[10pt]   \frac{(q\beta)^{-\frac{1+3w_\phi}{6w_\phi}}}{\left(4\pi\sqrt{3}\gamma\right)^{\frac{1+3w_\phi}{3(1+w_\phi)}}}\left(\frac{M_P}{M_{\rm in}}\right)^{-\frac{1+3w_\phi}{3(1+w_\phi)}}\left(\frac{\rho_{\rm end}}{M_P^4}\right)^{\frac{1+3w_\phi}{6(1+w_\phi)}}, 
    \end{cases}
    \end{align}
where the first expression applies for $\beta > \beta_{\rm *}$, and the second for $\beta < \beta_{\rm *}$. Here, $\beta_{\rm *}$ is the critical threshold that determines when the memory burden effect sets in. For $\beta > \beta_{\rm *}$, the burden effect occurs during the PBH-dominated phase. Otherwise, it takes place before PBHs come to dominate the energy density of the universe. The value of $\beta_{\rm *}$ can be estimated by equating the onset time of the burden effect, $
t_q = \frac{(1 - q^3)\,M_{\rm in}^3}{3\epsilon M_P^4} $ with the time at which early matter domination begins, $t_{\rm BH} \sim t_{\rm in}\,\beta^{-\frac{1+w_\phi}{2w_\phi}}$. This yields
\begin{eqnarray}
\beta_{\rm *} = \left[\frac{\epsilon}{4\pi\gamma(1+w_\phi)(1-q^3)\,S(M_{\rm in})}\right]^{\frac{2w_\phi}{1 + w_\phi}}\,.
\end{eqnarray}
\begin{figure}
    \centering
    \includegraphics[width=\linewidth]{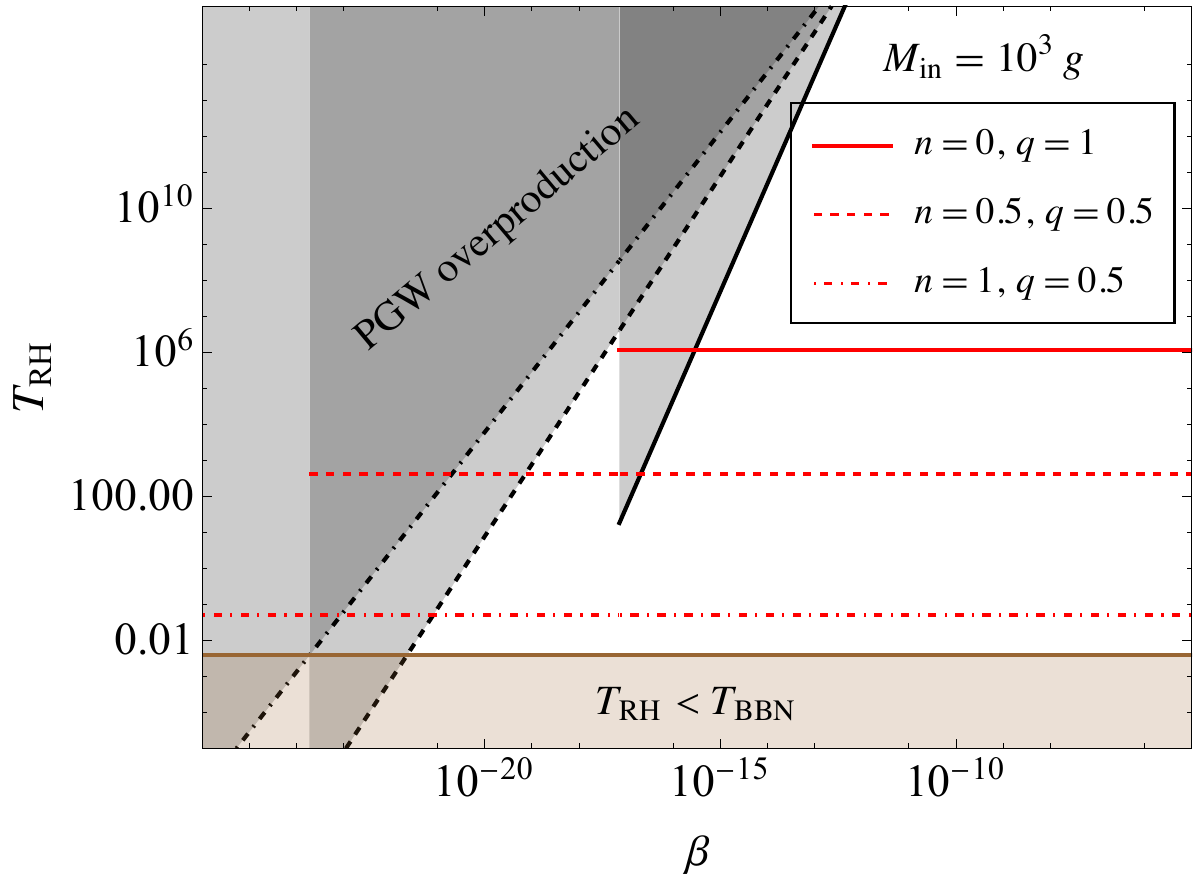}
    \caption{The restriction on the reheating temperature for $\beta > \beta_{\rm c}^n$ (i.e., when PBH domination occurs before evaporation) derived from the constraints on $\Delta N_{\rm eff}$ via PGWs as a function of $\beta$ is shown here. The plot includes standard Hawking evaporation ($n = 0,\,q = 1$) and two combinations of memory burden parameters, namely $(n = 1/2,\,q = 1/2)$ and $(n = 1,\,q = 1/2)$, depicted by the solid, dashed, and dot-dashed black curves, respectively. The black shaded region indicates where PGWs are overproduced, while the red curves denote the corresponding reheating temperatures for the PBH reheating scenario. The brown shaded region indicates reheating temperatures below the BBN energy scale, $T_{\rm RH} < T_{\rm BBN} \sim 4\, \rm MeV$ \cite{Kawasaki:1999na,Kawasaki:2000en,Hasegawa:2019jsa}, which are excluded by nucleosynthesis constraints.
    }\label{Fig: PGW-overproduction} 
\end{figure}
At this point, we note that throughout the analysis, we assume a scenario in which the memory burden effect begins during the PBH-dominated era. Although in some cases the starting point of burden effect may commence during the inflaton-dominated phase, determined by the equation of state $w_\phi$, the resulting correction enters through the $q$-factor, which affects the spectrum. Since this correction is typically small, we neglect such cases. Finally, the constraint on $T_{\rm RH}$ can be obtained by taking into account the overproduction of PGWs followed by the Eq. \eqref{eq: omegaineq}, leads to the following condition \footnote{Note that Eq. \eqref{eq: Trh-memory-cons} is not exactly accurate for the scenario $\beta < \beta_*$, as there will be a correction factor of $q^{\frac{3(3+2n)(1+w_\phi)}{6 - 2w_\phi + 4n(1+w_\phi)}}$. However, we ignore this correction since it does not significantly affect the constraint.}:
\begin{eqnarray}\label{eq: Trh-memory-cons}
\frac{T_{\rm RH}}{M_P} \leq 
\frac{2^{\frac{5(9 + 17w_\phi)+6n(5+9w_\phi)}{4(3 - w_\phi)+8n(1+w_\phi)}}}{3^{\frac{9 + w_\phi+2n(3+w_\phi)}{2(3 - w_\phi)+4n(1+w_\phi)}}}
\left( \frac{10^{-6}}{\Omega_{ GW}^{\rm rad}\, h^2} \right)^{\frac{3(1 + w_\phi)(3+2n)}{4(3 - w_\phi)+8n(1+w_\phi)}}\\
\left( \frac{\rho_{\rm end}}{M_P^4} \right)^{\frac{(3+2n)(1 - 3w_\phi)}{4(3 - w_\phi)+8n(1+w_\phi)}}
\nonumber \times \beta^{\frac{(3+2n)(1+w_{\phi})}{3-w_{\phi}+2n(1+w_\phi)}}\nonumber\\\left( \frac{\pi \gamma}{[\epsilon (3+2n)]^{\frac{1}{3+2n}}} \right)^{\frac{2w_\phi(3+2n)}{3 - w_\phi+2n(1+w_\phi)}} \frac{q^{\frac{(3+2n)(3w_\phi-1)}{6-2w_\phi+4n(1+w_\phi)}}}{\alpha^{1/4}}. \nonumber
\end{eqnarray}
Now let us examine how the constraint behaves in the PBH reheating scenario, where $\beta > \beta_{\rm c}^n$ (see, Fig.\ref{Fig: PGW-overproduction}), which is the focus of this study. We observe that for values of $\beta$ close to $\beta_{\rm c}^n$, the reheating temperature in PBH reheating scenario can potentially conflict with the $\Delta N_{\rm eff}$ bound. However, for larger $\beta$, corresponding to a prolonged period of PBH domination, this constraint becomes less stringent. This is because the PBH-dominated era effectively behaves as a matter-dominated phase, during which the gravitational wave background scales as $\Omega_{\rm GW} h^2 \propto k^{-2}$, suppressing the PGW contribution at high frequencies and evading the $\Delta N_{\rm eff}$ bound.

\section{PBH evaporation}

\label{Sec:pbhdecay}

\subsection{PBH reheating}

After inflation ends, the inflaton field typically undergoes coherent oscillations 
around the minimum of its potential. The effective background equation of state, $w_\phi$, depends on the shape of the potential, while the 
reheating duration is controlled by how rapidly the inflaton decays into Standard Model particles. In the standard picture, the Universe becomes radiation-dominated 
(RD) once the inflaton decays completely and its energy is transferred to relativistic species \cite{Garcia:2020eof,Haque:2020zco,Garcia:2020wiy,Haque:2022kez,Clery:2021bwz,Clery:2022wib}. 

In contrast to the standard reheating scenario, we consider a non-
standard post-inflationary history in which primordial black holes (PBHs) form during an inflaton-dominated epoch, and their subsequent evaporation gives rise to a radiation-dominated universe—a scenario referred to as \textit{PBH reheating} \cite{RiajulHaque:2023cqe}. A typical realization of this scenario occurs when the initial PBH abundance parameter $\beta$ exceeds a critical threshold $\beta_{\rm c}$. In this case, since PBHs behave as non-relativistic matter, they redshift more slowly than the inflaton background with an equation of state $w_\phi > 0$\footnote{Exact matter domination with $w = 0$ is excluded, as PBH formation in that regime is inefficient due to the long timescales required to form an apparent horizon \cite{Escriva:2020tak}.}, allowing PBHs to eventually dominate the energy density before they evaporate, thus initiating the radiation-dominated era \cite{Hidalgo:2011fj,Martin:2019nuw,Hooper:2019gtx,Hooper:2020evu,RiajulHaque:2023cqe}.
\footnote{Note that, $w_\phi < 1/3$, the inflaton is assumed to decay during the PBH-dominated phase, governed by its decay width. For $w_\phi > 1/3$, the inflaton must remain stable at least until the onset of PBH domination to ensure consistency with the cosmological framework.}. 

Even when PBHs do not dominate the energy density, i.e., for $\beta < \beta_c$, reheating can still occur through PBH evaporation. In such cases, unlike scenarios where PBHs dominate prior to evaporation, PBHs form and evaporate during the inflaton-dominated epoch without ever becoming the Universe’s main energy component. Reheating via PBH evaporation under these conditions is only possible if the inflaton behaves like a stiff fluid, meaning its equation of state satisfies $\omega > 1/3$, and if the inflaton decay or scattering rate remains sufficiently small. Specifically, the inequality $
\Gamma_\phi \rho_\phi (1 + \omega) < - \frac{\rho_{\rm BH}}{M_{\rm BH}} \frac{dM_{\rm BH}}{dt}
$ must be satisfied (see Refs. \cite{RiajulHaque:2023cqe,Haque:2023awl} for details). For the purposes of this study, we exclude this reheating scenario from our analysis.

The PBH evolution is governed by their formation mass $M_{\rm in}$ and initial energy fraction $\beta$. Assuming ultralight PBHs form in a $w_\phi$-dominated background, the evolution equations are:
\begin{align}
\frac{d\rho_\phi}{da} + 3(1+w_{\phi}) \frac{\rho_\phi}{a} &= -\frac{\Gamma_\phi \rho_\phi (1+w_{\phi})}{aH}\, , \\
\frac{d\rho_r}{da} + 4 \frac{\rho_r}{a} &\simeq -\frac{\rho_{\rm BH}}{M_{\rm BH}} \frac{d M_{\rm BH}}{da}\, , \\
\frac{d\rho_{\rm BH}}{da} + 3 \frac{\rho_{\rm BH}}{a} &= \frac{\rho_{\rm BH}}{M_{\rm BH}} \frac{d M_{\rm BH}}{da}\, , \\
\frac{dM_{\rm BH}}{da} &= -\epsilon \frac{M_p^4}{M_{\rm BH}^2} \frac{1}{aH}\, .
\end{align}

\noi
Here, $\epsilon = \left(\frac{27}{4}\right)\left(\frac{\pi}{480}\right) g_\ast(T_{\rm BH})$, where the factor $27/4$ originates from the greybody factor. This expression assumes that the greybody factor is evaluated in the geometrical-optics limit, as discussed in Refs. \cite{Arbey:2019mbc,Cheek:2021odj,Baldes:2020nuv}. In our case, we assumed that the radiation primarily comes from PBH evaporation, as inflaton decay is negligible ($\Gamma_\phi \ll H$).

Solving the mass loss equation gives:
\begin{equation}
M_{\rm BH} = M_{\rm in} \left(1 - \Gamma_{\rm BH}(t - t_{\rm in})\right)^{1/3}, \quad   \Gamma_{\rm BH} = \frac{3\epsilon M_P^4}{M_{\rm in}^3}\,.
\end{equation}

\noi
The PBH lifetime is $t_{\rm ev} = 1/\Gamma_{\rm BH}$, whereas the formation mass is
given by

\begin{equation}
M_{\rm in} = \frac{4\pi\gamma M_p^2}{H_{\rm in}}, \quad \mathrm{with}\quad \gamma = w_\phi^{3/2}\,,
\end{equation}

\noi
following the Carr-Hawking collapse efficiency \cite{Carr:1974nx} \footnote{A more accurate determination of the PBH formation mass depends on both the background equation of state $w_\phi$ and the shape of the primordial perturbations, as shown in Refs.~\cite{Musco:2012au, Musco:2008hv, Hawke:2002rf, Niemeyer:1997mt, Escriva:2021pmf, Escriva:2019nsa, Escriva:2020tak, Escriva:2021aeh}. However, a complete analytical treatment of the collapse efficiency parameter $\gamma$ in the post-inflationary regime is still lacking. Therefore, we adopt the conventional estimate proposed by Carr and Hawking.}. 

The standard Hawking evaporation neglects backreaction from emitted radiation, but Dvali et al.~~\cite{Dvali:2020wft,Dvali:2018xpy} showed this becomes relevant when the emitted quanta's energy is comparable to the black hole’s. This leads to the \textit{memory burden effect}~~\cite{Dvali:2020wft}, where systems with large information content resist decay until memory is transferred. Once about half the BH mass is lost, stored information slows further evaporation. This modifies PBH mass loss to~\cite{Alexandre:2024nuo}

\begin{eqnarray}
\frac{d M_{\rm BH}}{d a} = - \frac{\epsilon}{[S(M_{\rm BH})]^n}\frac{M_P^4}{M_{\rm BH}^2}\frac{1}{aH}\,,
\end{eqnarray}

\noi
with entropy $S = \frac{1}{2}(M_{\rm BH}/M_P)^2$ and memory parameter $n > 0$. The effect becomes significant when $M_{\rm BH} = q\,M_{\rm in}$, most strongly at $q = 1/2$~\cite{Alexandre:2024nuo,Thoss:2024hsr}. In this scenario, the critical value of $\beta$, above which PBHs come to dominate the Universe before fully evaporating, is significantly modified due to the memory burden effect, which prolongs the PBH lifetime. 

To estimate this critical value, we assume the burden effect sets in during the inflaton-dominated era. The time at which PBHs begin to dominate is then given by $
t_{\rm BH} \sim t_{\rm in} (q\beta)^{-(1 + w_\phi)/2w_\phi}$. On the other hand, the PBH lifetime is given by Eq.~\eqref{Eq:mass2}, i.e., $t_{\rm ev}^n = 1/\Gamma_{\rm BH}^n$. Equating these two timescales yields the modified critical value of $\beta$,

\begin{equation}
\beta_c^{ n} \sim \left[\left(\frac{M_P}{qM_{\rm in}}\right)^{2(n+1)}\frac{(3+2n)\epsilon}{2^{1-n}3\pi\gamma(1+w_\phi)}\right]^{\frac{2w_\phi}{1+w_\phi}} q^{-\frac{1+3w_\phi}{1+w_\phi}}\,. 
\label{Eq:betacburden}
\end{equation}

\noi
The above expression reduces to the standard semiclassical result, Eq.~\eqref{Eq:betamin}, in the limit $n = 0$ and $q = 1$ (corresponding to the case without memory burden). 

Let us now compute the reheating temperature for the case $\beta < \beta_{\rm c}$, although this scenario is not the main focus of the current work. As previously 
discussed, for $\beta < \beta_{\rm c}$ and a small inflaton decay rate $\Gamma_\phi \ll H$, the universe can be reheated via PBH evaporation, which occurs after a 
certain time, i.e., $a_{\rm RH} > a_{\rm ev}$. The reheating condition in this case is given by
\begin{equation}
\rho_{\rm RH} = \rho_{\rm BH}(a_{\rm ev})\left(\frac{a_{\rm ev}}{a_{\rm RH}}\right)^4 = \rho_\phi(a_{\rm in})\left(\frac{a_{\rm in}}{a_{\rm RH}}\right)^{3(1 + w_\phi)}\,,
\end{equation}

\noi
from which we find

\begin{align}
 &&T_{\rm RH}=\left(\frac{48}{\alpha}\right)^{\frac{1}{4}}\left[\frac{2^{n-1}(3+2n)\epsilon}{3(1+w_\phi)(\pi \gamma)^{3w_\phi}}\right]^{\frac{1}{2(1-3w_\phi)}} \beta^{\frac{3(1+w_\phi)}{4(3w_\phi-1)}} \nonumber \\
 &&\times \left(\frac{M_P}{M_{\rm in}}\right)^{\frac{3(1-w_\phi)+2n}{2(1-3w_\phi)}} q^{\frac{9+3w_\phi+4n}{4(3w_\phi-1)}}M_P\,,
\end{align}
where to find the above expression, we used $(a_{\rm ev}/a_{\rm RH})^4=(q\beta)^{4/(3w_\phi-1)}(a_{\rm in}/a_{\rm ev})^{12w_\phi/(1-3w_\phi)}$. On the other hand, for $\beta > \beta_{\rm c}^n$, the reheating temperature is determined by Eq.~\eqref{eq: tempmemory}, and can be expressed as

\begin{eqnarray}
T_{\rm RH} = \left(\frac{4}{3\alpha}\right)^{\frac{1}{4}}\sqrt{\frac{2^n(3+2n)\epsilon}{q^{3+2n}}}\left(\frac{M_P}{M_{\rm in}}\right)^{\frac{3}{2}+n}M_P\,.
\end{eqnarray}
\begin{figure}
    \centering
    \includegraphics[width=\linewidth]{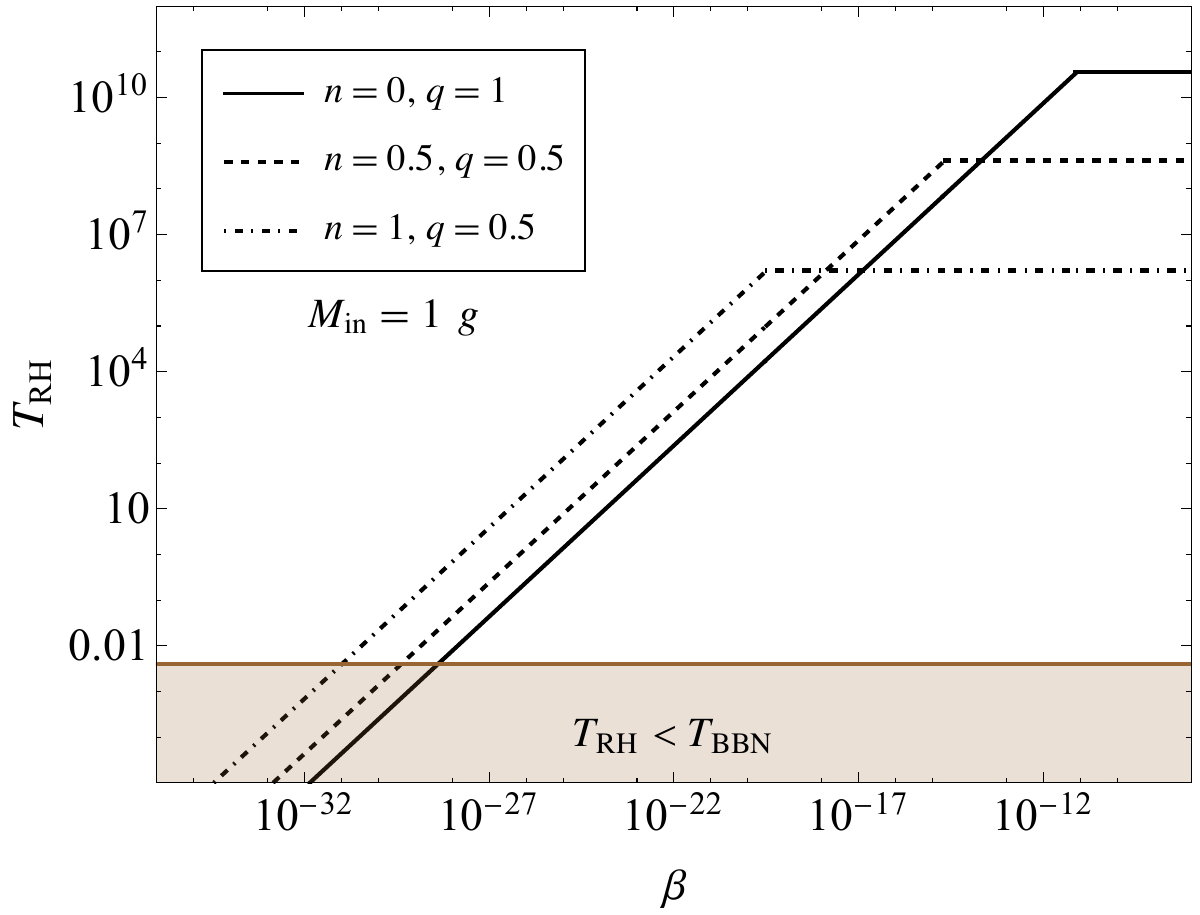}
    \caption{The reheating temperature as a function of $\beta$ is plotted here for standard Hawking evaporation ($n = 0,\, q = 1$) and for two different combinations of memory burden parameters: $(n = 1/2,\, q = 1/2)$ and $(n = 1,\, q = 1/2)$, represented respectively by the solid, dashed, and dot-dashed black curves.
    }\label{Fig: Trh}
\end{figure}

In Fig.\ref{Fig: Trh}, we show how the reheating temperature varies as a function of $\beta$. As expected, for values $\beta > \beta_{\rm c}^n$, which differ depending on the specific scenario, such as standard Hawking evaporation or different memory burden cases, we observe a $\beta$-independent plateau in the reheating temperature.

The maximum initial PBH mass that can fully evaporate before the onset of Big Bang Nucleosynthesis (BBN) can be estimated by requiring the PBH lifetime $t_{\rm ev}^n \sim 1\,\text{sec}$, and turns out as 
\begin{align}
&M_{\rm in} \leq \frac{M_P}{q} \left[2^n(3 + 2n)\epsilon\, t_{\rm BBN} M_P \right]^{\frac{1}{2n + 3}} \nonumber \\
&\sim \frac{1.1 \times 10^9\text{g}}{q} \left[2^n (3 + 2n)\right]^{\frac{1}{2n + 3}} \left(1.5 \times 10^{-29} \right)^{\frac{n}{2n + 3}}
\,.
\end{align}
The above equation implies that, under standard Hawking evaporation, the maximum allowed PBH formation mass is $M_{\rm in} \leq 1.6 \times 10^9 \,\text{g}$. However, when incorporating the memory burden effect with representative values $q = 1/2$ and $n = 1$, as adopted throughout our analysis, this bound becomes significantly tighter, yielding $M_{\rm in} \leq 6.1 \times 10^3 \,\text{g}$. Meanwhile, constraints on the tensor-to-scalar ratio from Planck+BICEP2/Keck ($r_{0.05}<0.036$)~\cite{BICEP:2021xfz,BICEP2:2015nss} imply an upper bound on the inflationary Hubble scale, $H_{\rm I} < 4.8 \times 10^{13}\,\text{GeV}$. This, in turn, sets a lower bound on the PBH formation mass $
M_{\rm in} \geq 0.5 \,{\rm g }\left( \frac{\gamma}{0.2} \right)$.

\subsection{Hawking evaporation}

The complete mass evolution of the PBH is described by a first phase of usual Hawking evaporation up to a time $t_{q}$ after which it will follow the burdened evolution described by Eq.~\eqref{Eq:mass_burden}. overall the mass evolution can be sumarized in the following form
\begin{align}
    M_{\mathrm{BH}}(t) &= M_{\mathrm{in}} \left( 1 - \gammabh^{(0)}(t - t_{\mathrm{in}}) \right)^{\frac{1}{3}}\theta(t_{q}-t) \nonumber\\&+ q\Min\left[1-\gammabh^{(n)}(t-\tq)\right]^{\frac{1}{3+2n}}\theta(t-t_{q})\,.
    \label{Eq:total_mass_burden}
\end{align}

From there, the temperature evolution and the instantaneous emissions will be 
modified according to the new PBH mass evolution. The temperature will be computed in the usual way, 
Eq.~\eqref{Eq:thawking} is still valid and will therefore exhibit the same "bouncing" 
behavior as the mass evolution. The more subtle point comes at the level of the 
instantaneous emission. The usual workaround for PBH evaporation is to get the 
generic instantaneous emission before integrating it with respect to the energy in 
order to get the proper mass evolution. Here we are working the other way around by 
imposing an extra $S^{-n}$ suppression at the level of the mass evolution. Luckily 
this term does not depend on the energy which result in a rather simple modification of 
the instantaneous emission
\begin{align}
    &\frac{d^2 N_{i,lm}}{dt \, dE} = (\theta(t_{q}-t)+\frac{\theta(t-t_{q})}{S^{n}})\frac{1}{2\pi} \, \frac{\Gamma_{s_i,lm}(E, M, x_j)}{e^{E/T_{\rm BH}} - (-1)^{2s_i}}\label{Eq:instantaneous_emission}\\
    &\sim(\theta(t_{q}-t)+\frac{\theta(t-t_{q})}{S^{n}})27\pi^{2}R_{S}^{2}\frac{g_{i}}{(2\pi)^{3}}\frac{E^{2}}{e^{E/T_{\rm BH}} - (-1)^{2s_i}}\,,
    \nonumber 
\end{align}

\noi
where we took the geometric optic limit in the last step. We also introduced here the greybody factor $\Gamma_{s_{i},lm}$ which encapsulates the reabsorbtion of the 
emitted particles by the PBH. Regarding the computations, we used the geometric 
optic limit but the figures take this factor into account since we used instantaneous emission curves from Black Hawk \cite{Arbey_Auffinger_2019,Arbey_Auffinger_2021} to produce our plots.

\subsection{The spectrum}
Let's now have a look at the shape of the PBH evaporation spectra in more details. The whole work is to properly integrate the contribution of the PBH population using the instantaneous emission relation given in Eq.~\eqref{Eq:instantaneous_emission}. The GW abundance can be reconstructed using 
\be
\frac{d \rho^{\mathrm{BH}}_{\mathrm{GW}}}{dt \, dk} = n_{\mathrm{BH}}(t) \, k \, \frac{dN_i}{dt \, dk}\,.
\ee
Since the memory burden affects only the latest stage of evaporation through the mass evolution, the evaporation can be expressed in the usual form \cite{Gross:2024wkl}
\be
\label{Eq:evaporation}
\frac{d \rho^{\mathrm{BH}}_{\mathrm{GW}}}{d \ln k_0}
= \frac{27 \, n^{\mathrm{BH}}_{\mathrm{in}} \, k_0^4}{64 \pi^3 M_P^4}
\frac{1}{S^{n}}\int_{a_{\mathrm{in}}}^{a_{\mathrm{ev}}}
\frac{M_{\mathrm{BH}}^2(a)}{\exp\left( \frac{k_0 M_{\mathrm{BH}}}{M_P^2} \frac{a_0}{a} \right) - 1}
\left( \frac{a_{\mathrm{in}}}{a} \right)^3
\frac{da}{H a}\, .
\ee
Here, the factor $M_{\rm{BH}}^{2}$ with Eq.~\eqref{Eq:total_mass_burden} splits
the integral in two contributions from each evaporation phase,
giving the 2-peaks structure of the spectra one can see in Fig.~\ref{Fig:masterplot} at the $D'_1$ and $D'_2$ spots. Eq.~\eqref{Eq:evaporation} then gives an estimation of the peak frequency if we approximate the mass of the PBH as constant before an abrupt evaporation,
\be
\label{Eq:peak_frequency}
f_{0,ev} \approx \frac{2.8}{2\pi}T_{\rm{BH}}\left(\frac{a_{ev}}{a_{0}}\right)\,,
\ee

\noi
which is the relation used in section~\ref{Sec:PBH evap}.
The evaporation scale factor $a_{ev}$ in Eq.\eqref{Eq:peak_frequency} can either be $a_{q}$ for $D'_1$ or $a_{RH}$ for $D'_2$. We can at this level grasp a fundamental difference between $D'_1$ and $D'_2$ since $D'_2$ is tied to the reheating temperature which is fixed by the last stage of PBH evaporation. This peak has therefore a fixed frequency which depends only on the mass while $D'_1$ feels on top, the redshift $a_{q}/a_{RH}$ which implies a dependence on the burden parameters.

\subsection{Value at the peak}
\label{Sec:pbhspectrum}

Starting from Eq.~\eqref{Eq:evaporation}, we can compute the main 
peak value for the PBH evaporation ($D_1$ and $D'_2$). The simplest way
is to express everything with respect to the reheating time 
while assuming the PBH mass as constant during the PBH domination era. We assume for simplicity that PBH domination occurs after the burden effect kicks in\footnote{The 
subtlety of the exact dynamic between the burdening and the domination can be 
technically taken into account but it only affects the tail of the spectra, which 
does not affect our results.}. Under this assumption, the PBH spectra can be expressed as \cite{Gross:2024wkl}

\beq
\left.\frac{d \rho}{d \ln k_0}\right|_{\beta>\beta_c}=
\frac{27 \sqrt{3}}{64 \pi^3}\frac{k_0^\frac52\sqrt{\rho_{\rm RH}}}{S^{n}\sqrt{q\Min}}\left(\frac{\aev}{a_0}\right)^\frac32\times I_0^{\beta_c}
\,,
\label{Eq:rhogwbhbetac}
\eeq
with

\beq
I_0^{\beta_c}=\int_{\frac{k_0 \Min}{M_P^2}\frac{a_0}{\aev}}^{\frac{k_0 \Min}{M_P^2}\frac{a_0}{\abh}}\frac{\sqrt{Y}}{e^Y-1}dY\,.
\eeq

\noi
Note that at this stage, our relations are generic. No other assumption has been made apart from the ones mentionned above.

The independence of the peak with respect to the mass of the PBH is also simple to 
demonstrate if we focus on the term that exhibits a mass contribution. The "hidden" 
mass factor are present in the entropy ($S$), the peak frequency ($k_{\rm{0}}$), 
the reheating energy density($\rho_{\rm{RH}}$) and the redshift between reheating and today ($a_{\rm{ev}}/a_{0}$). By massaging a bit those contribution we get
\beq
\left.\frac{d\rho}{dlnk_{0}}\right|_{\beta>\beta_{c}} \propto \frac{1}{S^{n}(q\Min)^{3}T_{\rm{RH}}^{2}}\propto (q\Min)^{0}.
\eeq
After collecting all the constant factor, the peak amplitude for the GW spectra can be expressed\footnote{note that here there is an extra $1/2$ difference compared to \cite{Gross:2024wkl} due to a slight difference in the convention of the reheating temperature}

\beq
\left.\Omega_{\rm{GW}}\right|_{\beta>\beta_{c}}^{\rm{peak}} = \frac{81(2.8)^{5/2}}{128\pi^{3}(3+2n)}\frac{\alpha}{\epsilon}\left(\frac{g_{0}}{g_{\rm{RH}}}\right)^{\frac{2}{3}}\frac{T_{0}^{4}}{3M^{2}_{P}H^{2}_{0}} I_{0}^{\beta_{c}}.
\eeq

Then, the peak value is overall the same as for the regular hawking evaporation ($n=0$, $q=1$, $ \Omega_{\rm{GW}}\sim 10^{-8}$). We only inherit a dependency on the burden parameter $n$ through the constant factor in front of the amplitude. 
For our benchmark point ($n=1,\,q=1/2$) the peak amplitude is then approximately the same up to an ${\cal O}(1)$ reduction factor, which fits with Fig.~\ref{Fig:masterplot}.

\section{Inflaton scattering}
\label{App:scattering}

After inflation, the Universe goes through a phase of
oscillation of the inflaton field $\phi$, following an average equation of state 
$P_\phi =w_\phi \rho_\phi$, with

\beq
w_\phi =\frac{k-2}{k+2}\,,
\label{Eq:wphi}
\eeq
for a potential \cite{Garcia:2020eof,Garcia:2020wiy}
\beq
V(\phi)=\lambda M_P^4 \left(\frac{\phi}{M_P}\right)^k\,.
\label{Eq:inflatonpotential}
\eeq

\noi
This potential can be seen as the limit $\phi \ll M_P$ of, for instance, 
a Starobinsky potential \cite{Starobinsky:1980te} of the form
\beq
V(\phi)=\frac34 m_\phi^2M_P^2\left(1-e^{-\sqrt{\frac23}\frac{\phi}{M_P}}\right)^2\,,
\eeq
or $\alpha$-attractor T-models~\cite{Kallosh:2013hoa}

\beq
V(\phi)=\lambda M_P^4\left[\sqrt{6} \tanh \left(\frac{\phi}{\sqrt{6}M_P}\right)\right]^k\,.
\eeq

The evolution of a dominant $\phi$-energy density is expressed in the Boltzmann equation

\beq
\dot \rho_\phi +3(1+w_\phi)H \rho_\phi\simeq 0\,,
\label{Eq:diffrhophi}
\eeq
which implies
\beq
\rho_\phi = V(\phi_0)=\rhoe\left(\frac{\ae}{a}\right)^{3(1+w_\phi)}=
\rhoe\left(\frac{\ae}{a}\right)^{\frac{6k}{k+2}}\,,
\label{Eq:rhophi}
\eeq
with $\rhoe$ being the inflaton energy density at the end of inflation.
As for the field $\phi$, it obeys the equation of motion 
\be
\label{Eq:eom_inflaton}
\Ddot{\phi}+3H\dot{\phi} +\frac{dV}{d\phi} \approx 0\,.
\ee
The solution to this equation can be parametrized as

\be
\phi(t) = \Phi(t)\underbrace{\sum_{\nu} \mathcal{P}_{\nu}(t)e^{-i\nu \omega t}}_{\mathcal{P}}\,,
\ee

\noi
where $\Phi$ is the envelope of the field and the $\mathcal{P}_{\nu}$ are the 
Fourier coefficients of the quasi periodic oscillatory function $\mathcal{P}$. Since 
we have a separation of scales between the fast oscillating scale (${\cal O}(m_{\phi})$) 
and the slower expansion of the universe (${\cal O}(H)$) we can decompose 
Eq.~\eqref{Eq:eom_inflaton} as \cite{Garcia:2020eof,Garcia:2020wiy,Garcia:2021iag}

\begin{align}
    \dot{\Phi} &= -\frac{6H}{n+2}\Phi\,,\\
    \dot{\mathcal{P}}^{2} &= \frac{2m_{\phi}^{2}}{k(k-1)}(1-\mathcal{P}_{}^{k})\,.
\end{align}

Solving this set of equations leads to a decaying envelope $\Phi$ and a oscillating frequency $\omega$ of the form

\begin{align}
    \Phi &\propto a^{-\frac{6}{k+2}},\\
    \omega &= m_\phi \sqrt{ \frac{ \pi k }{ 2(k - 1) }}\frac{ \Gamma\left( \frac{1}{2} + \frac{1}{k} \right) }{ \Gamma\left( \frac{1}{k} \right) }\,,
\end{align}

\noi
where the inflaton effective mass is defined with the usual convention, $m_{\phi} = \sqrt{V''(\Phi)}$. From this decomposition, the inflaton can be seen as an infinite sum of modes with energies $E = \nu\omega$ with a decreasing amplitude $\Phi_{\nu} = \Phi \mathcal{P}_{\nu}$.

Regarding the gravitational waves produced by the inflaton scattering, 
it is more convenient to develop $V(\phi)\sim\phi^{k}$ in harmonics since they 
appear in the vertices for the relevant diagrams. We denote them $(\mathcal{P}^{k})_{n}$. The rate of energy transfer between the inflaton and the 
gravitational waves can then be expressed as \cite{Choi:2024ilx}
\be
(1 + w_\phi)\Gamma_h \rho_\phi = \frac{\rho_\phi^2 \omega}{4\pi M_P^4} \Sigma^k\,,
\ee

\noi
with
\be
\Sigma^k = \sum_{\nu=1}^{\infty} \Sigma_\nu^k \equiv \sum_{\nu=1}^{\infty} \nu \left| (\mathcal{P}^k)_\nu \right|^2\,.
\ee

\noi
Finally, the gravitational waves spectra can be written, for $k\neq4$ \cite{Gross:2024wkl} 

\begin{align}
\label{Eq:inflaton_GW_k>6}
&\Omega_{GW}h^{2} =  \frac{h^{2}}{\rho_{c}^{0}}\frac{d \rho_{\mathrm{GW}}^{\phi}(a_0)}{df_0}
= \frac{h^{2}}{\rho_{c}^{0}}\frac{1}{a_0^4} \, \frac{d \left[ a^4 \rho_{\mathrm{GW}}^{\phi}(a) \right]}{df_0} \nonumber \\
&= \frac{h^{2}}{\rho_{c}^{0}}\frac{\sqrt{3} \, \rho_{\text{end}}^{3/2}}{4 M_P^3} \, \frac{k+2}{|k-4|}
\left( \frac{2\pi}{\omega_{\text{end}}} \right)^{\frac{3k - 3}{k - 4}}
\left( \frac{a_0}{a_{\text{end}}} \right)^{\frac{9}{k - 4}} \Sigma^k f_{0}^{\frac{4k-7}{k-4}}.
\end{align}

Note that Eq.~\eqref{Eq:inflaton_GW_k>6} does not hold for $k=4$, 
since in this case the inflaton energy scale $\omega$, redshifts as fast as the 
radiation it produces. This can be seen by computing the relation giving today's 
frequency of a graviton produced at a scale factor $a$,

\begin{align}
f_0(a) 
&= \frac{\omega(a)}{2\pi} \, \frac{a}{a_0} \nonumber\\
&= \frac{\gamma_k}{2\pi} M_P \left( \frac{\rho_{\text{end}}}{M_P^4} \right)^{\frac{k-2}{2k}} 
\left( \frac{a_{\text{end}}}{a} \right)^{\frac{3k - 6}{k + 2}} 
\left( \frac{a}{a_{\text{0}}} \right)\,.    
\end{align}

This relation grasps most of the physics needed to understand the shape of the spectrum. 
Indeed, if $k=4$ this relation clearly show a constant frequency, independent on $a$. This gives a a dirac shaped spectrum\footnote{In fact a forest of spectra peaked at $\nu \omega$.}. It is consistent with the idea that,
in this case, the Universe should not 
the reheating transition since the equation of state is kept unchanged before and after reheating. 
On the other hand, for $k>4$, the frequency behaves as some negative power law of 
the scale factor resulting in a spectra growing with $f_{0}$ since then later produced graviton have an overall smaller frequency.

\section{GW from density fluctuations}
\label{Sec:densityfluctuation}

\subsection{Generalities}

 Scalar induced gravitational waves have been shown to be an exciting possibility to 
 probe early Universe scenarios especially in the context of PBH reheating due to the sharp transition to radiation domination \cite{Papanikolaou:2020qtd,Domenech:2024wao,Balaji_2024,Bhaumik_2024,Paul:2025kdd,Inomata:2019ivs,Inomata_2019,Inomata:2020lmk}. The appendix presented here stands more as a 
 lightning introduction to get to our main result which is the analytic relation for 
 the scalar induced gravitational waves for a burdened PBH with an arbitrary inital 
 equation of state presented in Eq.~\eqref{EQ:UV_burdened} and \eqref{EQ:IR_burdened}.
 This extends the recent results of \cite{Balaji_2024,Domenech:2024wao},
 while also 
 shedding  light on the issue of non-linearity which sets the cutoff of the spectra 
 seen in figure \ref{Fig:masterplot}. For a more detailled analysis on the formalism, 
 we redirect the reader to a proper review\footnote{ Note that modifications of gravity or loop corrections can also have  impact on the 
background of gravitational waves \cite{Addazi:2024qcv, Frob:2025sfq}.} \cite{Domenech_2021}.
Starting from the perturbed metric

\begin{equation}
    ds^2 = (1 - 2\Phi) dt^2 - a^2(\tau) \left[ (1 + 2\Phi )\delta_{ij} + h_{ij}) dx^i dx^j \right],
\end{equation}

\noi
the tensor modes are sourced at second order by scalar fluctuations 
\cite{Domenech_2021,Domenech:2024wao,Papanikolaou:2020qtd}, whose equation of motion is 

\begin{equation}
    h''_{k,\lambda} + 2\mathcal{H} h'_{k,\lambda} + k^2 h_{k,\lambda} = \mathcal{S}_{k,\lambda}\,,
\end{equation}

\noi
where the source term ${\cal S}$ can be expressed as

\be
\mathcal{S}_{k,\lambda} = 4 \int \frac{d^3 q}{(2\pi)^3} \, e^{ij}_\lambda(\mathbf{k}) q_i q_j \, \Phi_{\mathbf{q}} \Phi_{|\mathbf{k}-\mathbf{q}|} f(\tau, q, |\mathbf{k}-\mathbf{q}|)\,,
\label{Eq:sklambda}
\ee

\noi
with 

\begin{align}
f(\tau, q, |\mathbf{k} - \mathbf{q}|) &= T_\Phi(q \tau) T_\Phi(|\mathbf{k} - \mathbf{q}| \tau)
\\&+ \frac{2}{3(1 + w)} \left( T_\Phi(q \tau) + \frac{T'_\Phi(q \tau)}{\mathcal{H}} \right)\nonumber\\&
\left( T_\Phi(|\mathbf{k} - \mathbf{q}| \tau) + \frac{T'_\Phi(|\mathbf{k} - \mathbf{q}| \tau)}{\mathcal{H}} \right)\,. \nonumber
\end{align}

Solving the evolution of the scalar potential accounting for the cosmological evolution, the gravitational wave signal during radiation domination can be expressed as \cite{Domenech_2021,Domenech:2024wao,Papanikolaou:2020qtd} 

\begin{equation}
    \label{Eq:Omega_GW_RD}
    \Omega_{\mathrm{GW}}(k) = \frac{k^2}{12 \mathcal{H}^2} \overline{\mathcal{P}_h(k, \tau)},
\end{equation}

with the tensorial power spectrum expressed as \cite{Espinosa_2018}
\begin{widetext}
\begin{align}
\label{Eq:tensorial power spectra}
    \bar{\mathcal{P}}_{h} &\approx \frac{c_{\rm{s}}^{4}}{8}\left(\frac{\mathcal{H}}{k}\right)^{2}\left(\frac{k}{k_{eva}}\right)^{8}
    \int_{0}^{\infty}dv \int_{|1-v|}^{1+v}du
    \left(4v^{2}-\left(1+v^{2}-u^{2}\right)^{2}\right)^{2}
    \bar{\mathcal{I}^{2}}_{\rm{osc}}
    P_{\Phi}(uk)P_{\Phi}(vk)S^{2}_{\Phi}(uk)S^{2}_{\Phi}(vk)
    \mathcal{T}_{\Phi}^{2}(uk)\mathcal{T}_{\Phi}^{2}(vk)\,.
\end{align}
\end{widetext}


\noi
The functions $S_{\Phi}$ are phenomenological suppression factors which take into account how the scalar potential decay during the transition to radiation domination and $\mathcal{P}_{\Phi}$ is the initial power spectra which in our case come from  \cite{Papanikolaou:2020qtd} :

\begin{equation}
    \mathcal{P}_{\Phi}(k) = 
\frac{2}{3\pi} \left(\frac{k}{k_{\rm UV}} \right)^3\,.
\label{Eq:isocurvature}
\end{equation}

\noi
In Eq.~(\ref{Eq:isocurvature}), $k_{\rm{UV}}$ is the scale related to PBH formation,
introduced in Eqs.~\eqref{Eq:f0uvnoburden} and \eqref{Eq:f0uvburden}.
From Eqs.~\eqref{Eq:Omega_GW_RD} and \eqref{Eq:tensorial power spectra}, we can
then redshift the GW abundance

\begin{equation}
    \Omega_{\mathrm{GW},0}(k) h^2 \approx 0.387 \left( \frac{g_*(T_{\mathrm{RH}})}{106.75} \right)^{-1/3} \Omega_{r,0} h^2 \Omega_{\mathrm{GW},\mathrm{RD}}(k)\,.
\end{equation}

\subsection{The spectrum}

Before diving in the details of the spectrum for the unburdened case, we should introduce specific ratio of scales that allow to simplify the expressions:

\begin{equation}
    \frac{k_{\rm{UV}}}{\kin}=\left(\frac{\beta}{\gamma}\right)^{\frac{1}{3}}\,,
\end{equation}
\begin{equation}
    \frac{k_{\rm{BH}}}{\kin} = \sqrt{2}\beta^{\frac{1+3w_{\phi}}{6w_{\phi}}}\,,
\end{equation}
\begin{equation}
    \frac{k_{\rm{RH}}}{\kin} = \left(\frac{\beta \epsilon}{2\pi\gamma}\frac{M_{\rm{P}}^{2}}{M_{\rm{in}}^{2}}\right)^{\frac{1}{3}}\,.
\end{equation}

From there, following \cite{Domenech:2024wao} the GW spectra today can be expressed as the sum of two contributions,

\begin{align}
    \label{EQ:UV_unburdened}
    \Omega_{\rm GW, res}^{\rm peak} = C^4(w)\frac{c_s^{7/3}(c_s^2-1)^2}{576 \times 6^{1/3}\pi}
&\left(\frac{k_{\rm BH}}{k_{\rm UV}}\right)^8
\\&\times\left(\frac{k_{\rm UV}}{k_{\rm RH}}\right)^{17/3} \left(\frac{k}{k_{\rm{UV}}}\right)^{\frac{11}{3}}, \nonumber
\end{align}

\begin{align}
    \label{EQ:IR_unburdened}
    \Omega_{\rm GW, IR}(k) = C^4(w) \frac{c_s^4 }{120 \pi ^2}\left(\frac{2}{3}\right)^{1/3} 
    &\left(\frac{k_{\rm BH}}{k_{\rm UV}}\right)^8  \\
    &\times \left(\frac{k_{\rm UV}}{k_{\rm RH}}\right)^{14/3}
    \left(\frac{k}{k_{\rm UV}}\right)\,. \nonumber
\end{align}

\noi
These two contributions are visible on the unburdened curve of Fig.~\ref{Fig:masterplot}, but even more clear through the break in the shaded spectrum in Fig.~\ref{fig:master_plot_cutoffs}, where the cutoff constraint has been released. Note also that we introduced $C(w)$, which is a numerical  constant 
determined by the matching of the scalar fluctuations transfer function 
between the inflaton domination and the PBH domination era. 
Its exact definition is given by \cite{Domenech:2024wao}

\be
C(w) = \frac{9}{20} \alpha_{\text{fit}}^{-\frac{1}{3w}} \left( 3 + \frac{1 - 3w}{1 + 3w} \right)^{-\frac{1}{3w}}\,,
\ee

\noi
where $\alpha_{\rm{fit}} \approx 0.135$ is a numerically determined constant.

\subsection{Burdening the spectrum}

Moving to the burdened spectrum, the overall physics is the same. 
The memory burden effect mainly affects {\it when} the PBHs decay 
and the duration of the evaporation. This induces 
modifications of the scale $k_{\rm{eva}}$ and the suppression factor.
Following \cite{Balaji_2024}, we get

\begin{equation}
    \frac{k_{\rm RH}}{k_{\rm in}} = \left[\frac{((3+2k)2^{n}\epsilon)^{2}}{36\pi^{2}\gamma^{2}}\left(\frac{M_{\rm in}}{M_{\rm P}}\right)^{2}\left(\frac{M_{\rm P}}{qM_{\rm in}}\right)^{6+4n}\beta^{2}\right]^{\frac{1}{6}}\,,
\end{equation}

and

\begin{equation}
    S_{\Phi , \rm{eva}}(k) = q\left(\frac{2}{\sqrt{3}}\frac{\kappa}{\sqrt{3\kappa-1}}\frac{k}{k_{\rm{RH}}}\right)^{-\frac{1}{3\kappa}}\,,
\end{equation}

\noi
where $\kappa$ is a function of $n$ introduced by convenience as

\begin{equation}
    \kappa = 1+\frac{2n}{3}\,.
\end{equation}

With this new variable, taking the unburdened limit $n=0$ and $q=1$ corresponds to
$\kappa = 1$, for which we can check that the above expressions converge to the unburdened case.

Finally, the last missing piece is the transfer function for the fluctuations which is

\be
\label{Eq:tansfert_function_Phi}
\mathcal{T}_{\Phi}(k) = \left( 5 + \frac{1}{C(w)} \left( \frac{k}{k_{\rm{BH}}} \right)^2 \right)^{-1}\,.
\ee
The shape of this transfer function causes the calculation to be divided 
between the IR and UV contributions. Modes respecting $k\gg k_{\rm{BH}}$ which 
are the ones which re-entered before PBH domination, are damped due to the 
pressure of the inflaton fluid that limit the growth of the fluctuations.
On the ohter hand, 
the ones entering during PBH domination ($k\ll k_{\rm{BH}}$) should not experience this effect.

With this in mind, using Eq.~\eqref{Eq:tensorial power spectra} and following appendix D and E of \cite{Domenech:2024wao}, the GW spectra during radiation domination can be expressed as in the previous case,
by the sum of two contributions

\begin{multline}
\label{EQ:UV_burdened}
    \Omega_{\rm{GW,res}}(k)= \frac{(2c_{\rm{s}})^{2\frac{3\kappa+2}{3\kappa}}C^{4}(w)q^{4}(c_{\rm{s}}^{2}-1)^{2}}{13824 c_{\rm{s}}\pi}\\
    \left(\kappa\sqrt{\frac{2}{3}}\sqrt{\frac{2}{3\kappa-1}}\right)^{-\frac{4}{3\kappa}}\left(\frac{k}{k_{\rm{UV}}}\right)^{5-\frac{4}{3\kappa}}\left(\frac{k_{\rm{UV}}}{k_{\rm{RH}}}\right)^{7-\frac{4}{3\kappa}}\left(\frac{k_{\rm{BH}}}{k_{\rm{UV}}}\right)^{8}\\
    \Theta_{\rm{UV}}(k),
\end{multline}

\noi
and

\begin{multline}
\label{EQ:IR_burdened}
    \Omega_{\rm{GW,IR2}} = \frac{C_{\rm{s}}^{4}C(w)^{4}q^{4}}{36\pi^{2}}\frac{\kappa}{9\kappa-4}\left(\sqrt{\frac{2}{3}}\frac{\kappa \sqrt{2}}{\sqrt{3\kappa-1}}\right)^{-\frac{4}{3\kappa}}\\ \left(\frac{k_{\rm{BH}}}{k_{\rm{RH}}}\right)^{8}\left(\frac{k_{\rm{RH}}}{k_{\rm{UV}}}\right)^{\frac{6\kappa+4}{3\kappa}}\left(\frac{k}{k_{\rm{UV}}}\right)\,.
\end{multline}

\noi
As for the previous equations the spectrum converges to 
Eqs.~\eqref{EQ:IR_unburdened} and \eqref{EQ:UV_unburdened} in the unburdened 
limit and reproduces the previous results existing in the literature \cite{Domenech:2024wao,Balaji_2024}.

\subsection{Non-linear cutoff}
As highlighted in \cite{Inomata:2020lmk,Domenech:2024wao,Balaji_2024}, during the period of PBH domination, fluctuations grow over time as described by the Poisson equation
\begin{equation}
    \Phi = \frac{3}{2}\left(\frac{\mathcal{H}}{k}\right)^{2}\delta_{\rm{PBH}}\,.
\end{equation}

\noindent 
The initial power spectrum of the fluctuations in Eq.~\eqref{Eq:isocurvature},
approaches unity when the momentum approaches $\sim k_{\rm{UV}}$. Consequently, the spectrum \eqref{EQ:UV_burdened} may overestimate GW production for momenta larger 
than a certain nonlinear scale $k_{\rm{NL}}$ that we need to estimate. This scale
is defined as the scale where $\delta_{\rm{PBH}} \sim 1$, based on the 
Poisson equation. Since the relevant quantity is the fluctuations during radiation 
domination, we must evolve the fluctuations until PBH evaporation, using the 
transfer function from Eq.~\eqref{Eq:tansfert_function_Phi}, and a suppression 
factor arising from the smooth nature of PBH decay \cite{Domenech:2024wao}. 
Collecting all the expressions, the non linear cutoff should then satisfy the condition

\begin{equation}
    k_{\rm NL,sup} \sim k_{\rm BH} \sqrt{\frac{5C(w)}{\frac{2C(w)}{3}\left(\frac{k_{\rm BH}}{k_{\rm RH}}\right)^{2}S_{\Phi} \Phi_{\rm init}-1}}\,.
\end{equation}

\noi
A good approximation for this scale comes from the pole inside the square root, while estimating $\Phi_{\rm init}\sim \sqrt{\mathcal{P}_{\Phi}}$
\begin{multline}
\label{Eq:NL_sup}
    k_{\rm NL,sup} \sim k_{\rm UV}\\\left(\frac{3}{2C(w)q}\sqrt{\frac{3\pi}{2}}\left(\frac{2\kappa}{\sqrt{9\kappa-3}}\right)^{\frac{1}{3\kappa}}\left(\frac{k_{\rm eva}}{k_{\rm eq}}\right)^{2} \left(\frac{k_{\rm UV}}{k_{\rm eva}}\right)^{\frac{1}{3\kappa}}\right)^{\frac{6\kappa}{9\kappa-2}}.
\end{multline}

\begin{figure}[H]
    \centering
    \includegraphics[width=\linewidth]{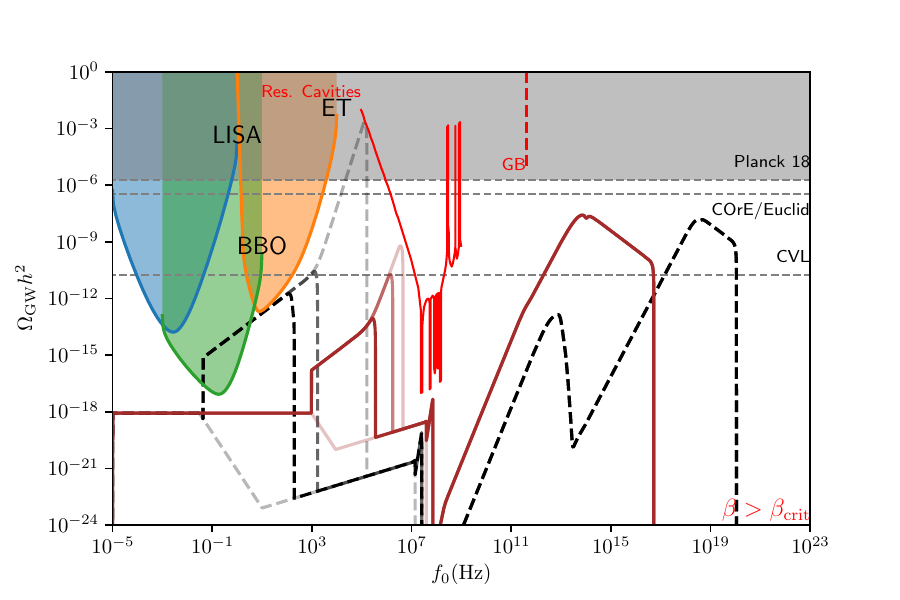}
    \caption{Fig.~\ref{Fig:masterplot} with different cutoffs. For each color, going from left to right, the most opaque is using $k_{\rm{cut}} = k_{\rm{NL}}$ from Eq.~\eqref{Eq:NL}, then the intermediate is using $k_{\rm{cut}} = k_{\rm{NL}}$ from Eq.~\eqref{Eq:NL_sup} and the most shaded is using $k_{\rm{cut}} = k_{\rm{UV}}$ from Eq.~\eqref{Eq:f0uvburden}.}
    \label{fig:master_plot_cutoffs}
\end{figure}

For completeness we can also compute the non-linear scale without the suppression factor which is a more conservative statement

\begin{equation}
\label{Eq:NL}
    k_{\rm{NL}} \sim \left(\sqrt{\frac{3\pi}{2}}\frac{3}{2C(w)}\right)^{\frac{2}{3}}\left(\frac{k_{\rm{eva}}}{k_{\rm{eq}}}\right)^{\frac{4}{3}}k_{\rm{UV}}\,.
\end{equation}
In all the figures presented above, we choosed to stay conservative using Eq.~\eqref{Eq:NL}. To let the reader appreciate the impact of the cutoff scale, we reproduce the Fig.~\ref{Fig:masterplot} while using different cutoffs with shaded curves in Fig.~\ref{fig:master_plot_cutoffs}.

\bibliography{ref}

\clearpage

\end{document}